\documentclass[useAMS,usenatbib,usegraphicx,twocolumn]{mn2e}

\usepackage{epsf}
\usepackage{comment}
\usepackage{fixltx2e}

\usepackage{amsmath}

\newcommand{\tablegap}{2mm}

\newcommand{\msun}{M_{\odot}}

\newcommand{\pc}{\, {\rm pc}}
\newcommand{\kpc}{\, {\rm kpc}}

\newcommand{\gyr}{\, {\rm Gyr}}

\def\gapprox{\;\rlap{\lower 3.0pt\hbox{$\sim$}}\raise 2.5pt\hbox{$>$}\;}
\def\lapprox{\;\rlap{\lower 3.1pt\hbox{$\sim$}}\raise 2.7pt\hbox{$<$}\;}

\newcommand{\figsizeFour}{5.5cm}

\newcommand{\figwidthDouble}{7.50cm}

\newcommand{\figsmall}{\figwidthDouble}

\newcommand{\reqOne}[1]{Equation~(\ref{#1})}
\newcommand{\reqTwo}[2]{Equations~(\ref{#1}) and~(\ref{#2})}
\newcommand{\reqNP}[1]{Equation~\ref{#1}}
\newcommand{\reqTwoNP}[2]{Equations~\ref{#1} and~\ref{#2}}

\newcommand{\be}{\begin{equation}}
\newcommand{\ee}{\end{equation}}

\newcommand{\ben}{\begin{enumerate}}
\newcommand{\een}{\end{enumerate}}

\begin{document}

\title[Velocity dispersion in globular clusters]{Application of three body stability to globular clusters II: observed velocity dispersions}
\author[Gareth F. Kennedy]
{Gareth F. Kennedy$^{1,2}$\thanks{Corresponding author email: gareth.f.kennedy@gmail.com}\\
      $^1$ National Astronomical Observatories of China, Chinese Academy of Sciences, Beijing 100012, China \\
      $^2$ Monash Centre for Astrophysics, Monash University, Clayton, Vic, Australia, 3800 
}
\maketitle

\begin{abstract}

The velocity dispersion profile in globular clusters (GCs) is explained here without having to rely on dark matter or a modification of Newtonian dynamics (MOND). The flattening of the velocity dispersion at large radii in certain Milky Way GCs, or lack thereof, is explained by recourse to the stability of the three-body problem in Newtonian dynamics. The previous paper in this series determined an analytical formula for the transition radius between stable and unstable orbits for a star in a globular cluster. This stability boundary is used here to predict where the velocity dispersion profile is expected to flatten in GCs, given known orbital parameters of the GC-galaxy orbit. Published observational data for the velocity dispersion as a function of radius of 15 Milky Way globular clusters with approximately known orbital parameters are used here.

We find that the stability boundary predicts flattening in the majority of clusters. While observational uncertainties in the orbital parameters prevent MOND from being ruled out entirely for some clusters it is not the preferred model in any cluster. Based on the results of this study we recommend further velocity dispersion observations and orbital determination for NGC 6171 and NGC 6341 as these are promising candidates for distinguishing Newtonian and MOND models. In particular, NGC 6171 may already be showing evidence of the chaotic diffusion of stars leading to flattening at the predicted stability boundary. 

\end{abstract}


\section{Introduction}

Globular clusters are thought to contain no dark matter, either due to their formation or to the evaporation of low mass dark matter particles \citep{BM2008}. In the case of subsequent evaporation, dark matter may still be present in the outer regions of some clusters. However, for most clusters, dark matter can safely be assumed to be absent from the cluster. This means that the cluster density and velocity dispersion profiles should fall off as predicted by a Newtonian dynamics model of a cluster in equilibrium.

The discovery of flattening in the velocity dispersion profile for the galactic globular clusters NGC 5139 ($\Omega$ Cen) and NGC 7078 (M 15) \citep{ScarpaWCen03}, where dark matter is thought not to exist, has generated a lot of excitement with some studies claiming that this is direct evidence for a breakdown of Newton's laws at low accelerations. Explanations for the observed deviation broadly fit into three categories; tidal interactions, dark matter or a modified gravity theory (e.g. MOND). All of these theories can produce a flattening of the velocity dispersion profile for large radii, in the case of MOND this occurs at the critical acceleration of $1.2 \times 10^{-10}$ m/s$^2$ \citep{Milgrom1983}.

Distant GCs are considered as a test of different gravity theories \citep[e.g.][]{BGK2005,MT2008}, provided that the velocity dispersion profile is sufficiently resolved by being based on 30-80 stars in each cluster \citep{HBK2011}. At present the statistics are not quite sufficient to rule out MOND, for example in the case of Pal 14 \citep{GFAK2010}.

A distant cluster that has received a lot of recent attention in the literature is NGC 2419. This cluster is located approximately 90 kpc from the galaxy and as such, tidal interactions with the galaxy are expected to be negligible. In addition it is a very luminous cluster and contains many bright giants suitable for spectroscopic analysis, greatly aiding the accuracy of the velocity dispersion profile.

The velocity dispersion profile of NGC 2419 can only be fit using the MOND acceleration cut-off if there is substantial radial anisotropy in the cluster \citep{SN2010}. The effect of radial anisotropy was found to be stronger in a MOND cluster profile than their equivalent Newtonian systems \citep{NCL2011}. While there is some radial anisotropy due to rotation for NGC 2419 \citep{IbataEtAl2013}, \citet{BCHRMDS2009} found that there is not enough to reconcile the velocity dispersion with MOND. NGC 2419 is most likely to be a remnant of a much larger system as evidenced dynamically \citep{BCHRMDS2009} and chemically \citep{ChemNGC2419}, which complicates the picture but still fails to reconcile the observations to any MOND model. A more recent and comprehensive comparison between Newtonian and MOND models to an increased data set for NGC 2419 by \citet{IbataEtAl2011} found that the best MOND fit was far worse than the best Newtonian Michie model.

For closer GCs, recent observational studies have found that MOND or dark matter is not required in many GCs. For example \citet{LaneEtAl2011} (and references therein) found that of the 10 GCs studied, the velocity dispersion profiles of all except NGC 6121 (M 4) could be well fit using a Plummer sphere model. They found that NGC 6121 had a mass of twice the literature values due to tidal heating increasing the velocity dispersion in the outer regions. The velocity distribution of this cluster was also complicated by the clear signature of cluster rotation found in the observations.

The premise of this paper is that the velocity dispersion profile is expected to flatten out in the region where orbits become unstable. This effect is distinct to tidal interactions, such as tidal shocking, as it is based on properties of the general point-mass three-body problem. The method used for determining the stability boundary is described in detail in the previous paper in this series \citep[][hereafter Paper I]{PaperI} which was based on the stability of the general three-body problem as derived in \citet{RoIoA} and \citet{RoNew}. This instability occurs in the Newtonian dynamics regime without the need for any modifications to the theory of gravity or additional dark matter. The determination of the radius associated with the stability boundary for a Plummer sphere is summarised in Section~\ref{SBPlummer}. This section also covers the velocity dispersion as a function of radius for Newtonian dynamics and the radius where this profile is predicted to break down according to MOND. A simulated cluster model is used in Section~\ref{SimCluster} to show that the velocity dispersion profile does indeed flatten where stars exist on unstable orbits. 

Section~\ref{AppMWGCS} examines the orbits of 15 GCs from the Milky Way globular cluster system where observational data is available for the velocity dispersion as a function of radius and on the velocity of the cluster. After determining the GC-galaxy orbits, the predicted radii for the stability boundary and MOND models are compared to the velocity dispersion data. This section also examines the observational uncertainties in the orbits and cluster masses that affect the analysis. Statistical comparisons between the different models are done in Section~\ref{Comparison} and discussed in Section~\ref{IndivGCs} along with complications to the model comparisons. Conclusions are drawn in Section~\ref{Conc}, in particular that modified gravity theories are not needed to explain the observations.

\section{Stability boundary in GCs}
\label{SBPlummer}

\subsection{Simple Plummer sphere model}

\label{SimplePlummerCluster}

The previous paper in this series outlined a simple method for determining the transition between stable interior and unstable exterior orbits in a GC. An unstable system is taken to mean that a star is on an orbit that makes the total system unstable to the escape of one of the bodies. In the context of GCs this is equivalent to the eventual escape of a star from the potential well of the cluster. Paper I used this stability boundary to estimate the tidal radius for a cluster given its galactic orbital parameters of perigalacticon ($R_p$) and eccentricity ($e$). This radius was derived assuming point masses for the galaxy, the cluster centre and a cluster star. A Plummer potential was used to provide an eccentricity distribution for the stars orbits, which was averaged over to determine the stability boundary. In this section we give an overview of the results from the previous paper where they are relevant to predicting the flattening of the velocity dispersion in GCs.

A globular cluster is modelled using a Plummer sphere with gravitational potential given by \citep{BinneyTremaine1987}
\begin{equation}
\Phi=\frac{-GM_{C}}{\sqrt{r_{h}^{2}+r^{2}}}
\label{PlummerPotentialUS}
\end{equation}
where $G$ is the gravitational constant, $M_{C}$ is the mass of the cluster, $r$ is the radial distance, and $r_{h}$ is the observable half mass radius of the cluster for a projected velocity dispersion.

\citet{Dejonghe1987} gives the velocity dispersion as a function of cluster radius, $r$, for a Plummer sphere to be
\begin{equation}
\sigma^2(R) = \frac{\sigma_0^2}{\sqrt{1+\frac{r^2}{r_{h}^2}}}
\label{SigmaR}
\end{equation}
which is referred to as the equilibrium velocity dispersion and the central velocity dispersion ($\sigma_0$) is related to the cluster mass $M_C$ by
\begin{equation}
M_C = \frac{64 \sigma_0^2 r_{h}}{3 \pi G}
\label{MCSigma}
\end{equation}
where $G$ has its usual value. There are a number of implicit assumptions in \reqOne{MCSigma}; firstly that the cluster is composed of equal mass stars from a single population, secondly that cluster rotation is negligible and finally that the binary fraction is effectively zero. 

Multiple stellar populations and rotation in globular clusters make getting the mass directly from the central velocity dispersion fraught with dangers \citep{Bekki2010}. The review by \cite{GCB2012} note that all clusters where chemical observational data exists show evidence of multiple populations. For our analysis this will lead to additional uncertainty in the total mass of the cluster. The assumption of equal mass stars is not expected to change our results since the mass ratio between the star and the cluster is approximately $10^{-5}$, so an additional order unity change due to the mass function does not strongly affect the stability boundary (see Paper I).

The effect of rotation on the large scale dynamical evolution of a cluster was examined using Fokker-Planck models by \cite{EinselSpurzem1999} who found that the central velocity dispersion changed by up to 10\% compared to non-rotating clusters. This results was also confirmed using N-body simulations \citep{BaumgardtHutHeggie2002}. This can lead to poor fits to the total cluster mass and to the validity of the equilibrium model for the velocity dispersion. The measured rotation for each cluster is examined in Section~\ref{Comparison} as a potential source of error in the velocity dispersion analysis.

The final assumption made was that binaries are unimportant. \cite{ThijsRichard2009} found that binary systems are only important in cases where $\sigma \lapprox 1$km/s and negligible for $\sigma \gapprox 10$ km/s. The clusters of interest in this paper are all in the range of $\sigma \gapprox 1$ km/s in the central regions where the binary fraction is expected to be maximum. As we are interested in the outer regions of the cluster any effect of stellar binaries will be neglected from herein.

As shown in Section~\ref{AppMWGCS}, the cluster masses given in the literature have a large spread between studies, as they often used different methods and mass to light ratios. For this reason we use the mass derived from the central velocity dispersion as a consistent method for all GCs. In addition to consistency, this method provides a range of $M_C$ values that provide adequate fits to the inner velocity dispersion measurements, effectively combining all of the above uncertainties due to our assumptions into a single quantifiable uncertainty.

\subsection{Globular cluster radii}
\label{GCRadPred}

Throughout this paper there are three radii of key importance for each GC. These are the radius where the acceleration acting on a star due to the cluster potential goes beneath the MOND limit of $a_0 = 1.2 \times 10^{-10}$ m/s$^2$ ($r_{m}$), the tidal radius ($r_t$) and the radius of the transition between stable and unstable orbits ($r_{c}$). Firstly the MOND radius is found by iterating over the acceleration $a_0$ and solving for $r_{m}$, which for a Plummer sphere is
\begin{equation}
a_0 = \frac{G M_C r_{m}}{\left( r_{h}^2 + r_{m}^2 \right)^{3/2}}
\label{ra0}
\end{equation}
given the known values of the cluster mass ($M_C$) and the half-mass radius ($r_{h}$).

For the remaining radii we adopt a functional form for the tidal radius that is separable into a mass component and orbital eccentricity of the cluster-galaxy orbit 
\begin{equation}
r_t = R_p \left( \frac{M_C}{M_G} \right)^{1/3} f(e)
\label{rtidal}
\end{equation}
where $M_G$ is the effective point mass galaxy, $e$ is the eccentricity of the cluster orbit around the galaxy, and $R_p$ is the distance of closest approach to the galaxy, referred to as perigalacticon. This form of the tidal radius also allows direct comparison between the stability boundary and other tidal radii estimates from the literature.

The most commonly used tidal radius estimate is for a star on a radial orbit and using point mass potentials for the cluster and galaxy. The eccentricity dependence, $f(e)$, of the tidal radius is given by the classical \citet{KingI} result
\begin{equation}
f(e) = k \left( 3 + e \right)^{-1/3}
\label{rtidal_King}
\end{equation}
where the constant $k \sim 0.7$ was introduced by \citet{Keenan1981} to better fit observations of the galactic globular clusters. The tidal radius ($r_t$) will be used to denote the maximum theoretical tidal radius of a GC by using \reqTwo{rtidal}{rtidal_King}.

The final radius of interest is the stability boundary that was determined in Paper I using the eccentricity distribution for stars in an isolated cluster as modelled by a Plummer sphere of potential given by \reqOne{PlummerPotentialUS}. The transition between stable interior orbits and unstable exterior orbits was not a single value; rather it was a range of radii that depended on the eccentricity of the cluster-galaxy orbit. To describe this transition three values were used, the minimum and maximum values and an indicative value referred to as the chaos radius and denoted by $r_{c}$.

\begin{figure}
\begin{centering}\resizebox{\figsmall}{!}{\includegraphics[angle=270]{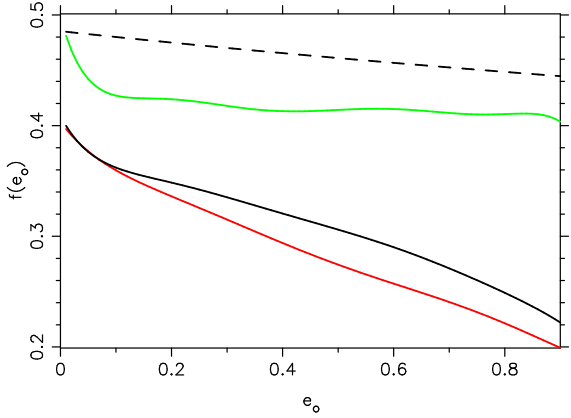}} 
\par\end{centering}
\caption
{The eccentricity dependence for the transition from stable inner orbits to unstable exterior orbits. The form of the eccentricity dependence is given by \reqOne{rtidal_MSC} with coefficients given in Table~\ref{Coefficients} and the corresponding radius is given by \reqOne{rtidal}. The indicative value for the stability boundary ($r_{c}$) is shown as a black curve, while the minimum and maximum extents of the partially stable region are the red and green curves respectively. The King radius (Equation~\ref{rtidal_King}) is shown as the dashed black curve for comparison.}
\label{fig:rtidal}
\end{figure}

Using the same form for the tidal radius as \reqOne{rtidal} then the eccentricity dependence for all three values describing the stability boundary can be fit using a Taylor series in $e$ of the form
\begin{equation}
f(e) = \exp \left[ \sum_{i=0}^{N=7} a_i e^i \right]
\label{rtidal_MSC}
\end{equation}
where the coefficients are given in Table~\ref{Coefficients} for the indicative stability boundary radius ($r_{c}$) and the minimum and maximum extents of the partially stable region. For radii $r< r_{min}$ all stars are expected to be on stable orbits, whereas for $r > r_{max}$ they are expected to be on unstable orbits. As discussed in detail in Paper I, any star on an unstable orbit will undergo chaotic diffusion until it escapes the cluster. The eccentricity dependence for $r_{min/max}$ and $r_{c}$ as a function of the eccentricity of the cluster-galaxy orbit are shown in Figure~\ref{fig:rtidal}. 

\begin{table}
 \caption{Coefficients for the polynomial fit to $r_{c}$ and the minimum and maximum extents of the marginally stable zone.}
 \label{Coefficients}
\begin{centering}
 \begin{tabular}{|l|l|l|l|}
\hline 
 & $f_{chaos}$ & $f_{min}$ & $f_{max}$ \\
\hline 
$a_0$	&	-0.89462	&	-0.907914	&	-0.696568	\\
$a_1$	&	-2.36353	&	-1.67172	&	-3.83438	\\
$a_2$	&	17.6103	&	7.2727	&	36.6515	\\
$a_3$	&	-77.3899	&	-23.8429	&	-177.338	\\
$a_4$	&	185.385	&	36.8088	&	461.984	\\
$a_5$	&	-247.308	&	-22.2994	&	-657.971	\\
$a_6$	&	171.836	&	-1.80097	&	482.319	\\
$a_7$	&	-48.6078	&	4.73382	&	-142.327	\\
\hline
 \end{tabular}
\end{centering}

 \medskip
For radii $f(e)< f_{min}$ all stars are expected to be on stable orbits, whereas for $f (e)> f_{max}$ they are expected to be unstable orbits. The resultant functions of the eccentricity of the cluster-galaxy orbit are shown in Figure~\ref{fig:rtidal}. For details on how these coefficients are determined see Paper I.
\end{table}

The functional dependence on the eccentricity was derived using a distribution of stellar orbital eccentricities and was effectively independent of the cluster mass. The assumption of a point mass potential for the galaxy for distant cluster orbits is justified for distant GCs since the mass inside 6.4 kpc is consistent with that of a point mass of roughly $10^{11} \msun$, based on the orbits of NGC 2419 and NGC 7006 \citep{Bellazzini2004}. However almost all of the GC orbits examined here have perigalacticon distances ($R_p$) inside 6 kpc. The procedure adopted in Section~\ref{GCOrbits} determines an effective Keplerian orbit consisting of $R_p$ and $R_a$ by averaging the minimum and maximum distances to the galactic centre over the most recent 10 passages. 
Thus the orbital parameters are themselves an approximation of a Keplerian orbit which is in turn an approximation to a more complicated orbit. In this context the error in the GC-galaxy orbit due to modelling the galaxy as a point mass is negligible. The overall effect of the uncertainties in the GC-galaxy orbit is discussed in Section~\ref{obsuncert}.

\subsection{Timescale correction to chaos radius}
\label{theory:time}

In paper I the chaotic diffusion process for a star on an unstable orbit was found to take approximately 10 GC-galaxy orbits. During this time chaotic diffusion can be suppressed if energy is transferred to or from the star-cluster-galaxy system. The most probable mechanism for this to occur is if the star encounters another star during the chaotic diffusion timescale.

A star will have its motion significantly altered by distant encounters with another star on the two-body relaxation timescale. To determine the two-body relaxation time near the stability boundary requires the two-body relaxation time at the half-mass radius given by \citep{BinneyTremaine1987} 
\begin{equation}
t_{rh} = \frac{0.78 \gyr}{\ln(\gamma N)} \left( \frac{1 \msun}{m} \right) \left( \frac{M_C}{10^5 \msun} \right)^{1/2}
\left( \frac{r_h}{\pc} \right)^{3/2}
\label{trxhalf}
\end{equation}
where $N$ is the number of stars in the system and $m$ is the mean stellar mass. By describing the GC as a Plummer sphere then the density profile in the outer regions is $\rho \propto r^{-5}$ and the velocity dispersion is $v \propto r^{-1/2}$ then the radial dependence of the relaxation timescale is
\begin{equation}
T_{rx}(r) \propto \frac{v^3}{\rho} \propto r^{7/2}.
\label{trxraddep}
\end{equation}
Simplifying the GC-galaxy orbit by a Keplerian orbit and taking $\gamma = 0.4$, $m=1 \msun$ and $N = M_C/\msun$, then the ratio of the GC-galaxy orbital period to the relaxation timescale is
\begin{eqnarray}
\nonumber
\frac{T_o}{T_{rx}} 
& \approx & 
(12.01) 
\left( \frac{M_C}{10^5 \msun} \right)^{-1} 
\left( \frac{r_h}{\pc} \right) ^{-3/2} 
\left( \frac{R_p}{\kpc} \right) ^{3/2} \\
 & \times & 
\left( 1-e_o \right) ^{-3/2}
\left( \frac{r}{r_{h}} \right)^{-7/2}.
\label{gc_relaxtime}
\end{eqnarray}

The radially dependent relaxation timescale can be used with the assumption that the timescale for a star to random walk out of the cluster takes 10 GC-galaxy orbits. Thus the chaos radius will satisfy this condition if $T_{rx}(r_c) \lapprox 10 T_o$ where $T_o$ is the radial period of the GC-galaxy orbit. In many clusters this relation will not be satisfied and so an effective chaos radius $r_c^*$ is sought such that $r_c^* > r_c$ and $T_{rx}(r_c^*) = 10 T_o$. The corrected chaos radius is then given by
\[
r_c^* = 
\begin{cases}
r_c, & \text{if } T_{rx}(r_c) > 10 T_o \\
r_c \left( \frac{10 T_o}{T_{rx}(r_c)} \right)^{2/7}, & \text{if } T_{rx}(r_c) < 10 T_o \text{ and } r_c^* < r_t \\
r_t, & \text{if } T_{rx}(r_c) < 10 T_o \text{ and } r_c^* > r_t \\
\end{cases}
\]
\label{CorrectedRc}
where $T_o/T_{rx}(r_c)$ is determined by \reqOne{gc_relaxtime} and $r_t$ is the standard King tidal radius defined by \reqTwo{rtidal}{rtidal_King}. This corrected radius forms the basis for using the stability boundary to predict where the velocity dispersion will flatten in the Milky Way globular clusters in Section~\ref{AppMWGCS}.

Note that neither the tidal radius nor the stability boundary act as an instant remover of stars, as pointed out by \citet{FH2000} and more recently investigated numerically by \cite{KupperEtAl2010}. They found for GCs on circular orbits that the escape timescales for stars beyond the tidal radii could be long enough to allow some stars to stay in this region indefinitely. This means that the predicted tidal radius will be a lower limit as stars outside this can still be close to the GC while remaining formally unbound. Stars on such unstable orbits will be observable inside the King radius, but are expected to have very different velocities compared to an equilibrium cluster. The effect of stars on these orbits on the observable velocity dispersion and surface brightness profiles are discussed in the next section. 

\label{cloud}

\section{Observational consequences of unstable orbits}

\label{SimCluster}

This section aims to determine what effect, if any, stars on chaotic orbits will have on observable GC profiles. To test for such an effect numerical simulations are made of the star-cluster-galaxy system for two cases; one stable and the other unstable.

The star-cluster-galaxy system is modelled by a three-body point mass system with $m_1 = 1 \msun$, $m_2 = M_C = 10^6 \msun$ and $m_3 = M_G = 10^{11} \msun$. The two cases both have a period ratio of $T_o/T_i = 15$ where $T_o$ is the cluster-galaxy orbital period and $T_i$ is the star-cluster period. To best illustrate the difference between stable and unstable orbits the system initially has outer eccentricity ($e_o = 0.5$), relative inclination ($I = 30\deg$), both orbits begin at apocentre and all other Euler angles are set to zero. The only difference between the two cases is the star-cluster orbital eccentricity; one has $e_i = 0.0$ and is expected to be stable, the other has $e_i = 0.3$ and is expected to be unstable. All particles are numerically integrated for a total of 10 cluster-galaxy orbits with a Bulirsch-Stoer integrator \citep{NR86} that conserves energy and angular momentum.

\begin{figure}
\begin{centering}\resizebox{\figsmall}{!}{\includegraphics[angle=270]{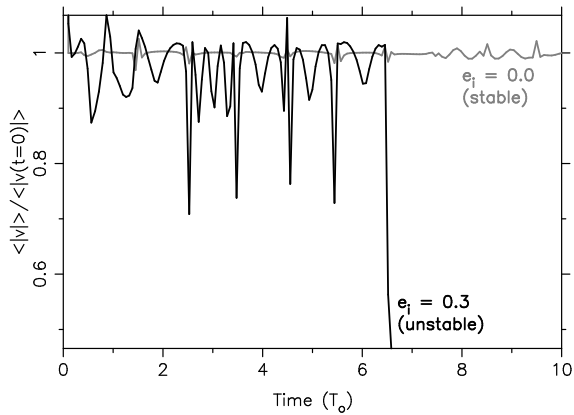}} 
\par\end{centering}
\caption
{Variation in the orbit averaged velocity magnitude of a star orbiting in a cluster for a stable orbit ($e_i=0.0$) and an unstable orbit ($e_i=0.3$). Three-body integrations are done using $T_o/T_i = 15$, $e_o = 0.5$ and $I = 30\deg$ with $m_1 = 1 \msun$, $m_2 = M_C = 10^6 \msun$ and $m_3 = M_G = 10^{11} \msun$. Similar plots were also obtained for each velocity component independently.}
\label{fig:VD_Sim} 
\end{figure}

Figure~\ref{fig:VD_Sim} shows the variation in the velocity magnitude of the star, averaged over the star-cluster orbital period, as a function of time. Stable orbits are not expected to vary significantly over time, except minor changes due to secular evolution and in particular Kozai cycles. This is true for eccentric orbits as well as circular orbits. Minor variations are seen for the $e_i = 0.0$ case in Figure~\ref{fig:VD_Sim} where the averaged velocity magnitude changes by less than 5\% compared to the initial velocity magnitude, $<$$|$$v(t=0)$$|$$>$. In contrast the unstable orbit ($e_i = 0.3$) exhibits large fluctuations (up to $\sim$ 30\%) before the star eventually escapes the cluster at $\sim 6.5$ GC-galaxy orbits ($T_o$). This particular unstable orbit is fairly long-lived, which is precisely the type of orbit that will survive long enough to contribute to the observable velocity dispersion profile. 

Since the velocity dispersion ($\sigma$) is the variation of velocities from the mean value, then a single star on an unstable can easily affect the $\sigma$ value in a given radius bin. By contrast the surface brightness is based on the total star count in the same radial bin, so the addition or subtraction of a single star will not significantly change the surface brightness. Therefore it is predicted that stars on unstable orbits at large distances from the cluster centre lead to a different velocity profile than an equilibrium model (such as \reqNP{SigmaR}).

To check that no effect is seen in the surface brightness the most recent surface brightness profiles for the GCs listed in Section~\ref{AppMWGCS} were examined \citep[data taken from][]{TragerEtAl1995,McLaughlinVDMarel2005,BianchiniEtAl2013,DiCeccoEtAl2013,MiocchiEtAl2013}. All comparisons for the chaos and tidal radii (values taken from Table~\ref{TblRadii}) to observed surface brightness profiles are based on cluster distances from \cite{Harris1996} (2010 edition). To summarise; deviations were seen beyond the tidal radius for NGC 288, NGC 1904, NGC 6218 and NGC 6341 as one would expect, possible deviations were seen beyond the stability radius in the rotating clusters NGC 5139, NGC 6121, NGC 6809 and NGC 7078, a very weak deviation was seen in NGC 5024 and NGC 6171, and no data was available at distant enough radii for NGC 1851, NGC 6656, NGC 6752 and NGC 7099. We conclude that the surface brightness data is too inconclusive to be used to detect deviations from the equilibrium model. Especially considering the large uncertainties in the GCs orbital parameters (see below).

For the remainder of this paper potential deviations in the surface brightness due to stars on unstable orbits are ignored and the focus will be on deviations in the velocity dispersion profile.

\section{Application to Milky Way GC system}

\label{AppMWGCS}

\subsection{Determination of GC-galaxy orbits}

\label{GCOrbits}

Before comparing the velocity dispersion profiles of real clusters to the radii estimates in Section~\ref{SBPlummer} the orbital parameters of the cluster-galaxy orbit are required. In this section the galactic orbits of 15 GCs are determined based on the observed velocities for each cluster and integrating backwards in time through a realistic galactic potential. The perigalacticon ($R_p$) and eccentricity ($e$) are then obtained from the integrated orbits for each GC.

Recent velocity and distance values for each cluster (references in Table~\ref{TblObs}) are used in conjunction with the galactic potential used by \citet{FellhauerEtAl2007}. This gravitational potential consists of a Miyamoto-Nagai potential \citep{MiyamotoNagai1975} combined with a logarithmic potential. The total galactic potential $\Phi$ is given as a sum of the galactic halo $\Phi_h$, disc $\Phi_d$ and bulge $\Phi_b$ potentials by \citep{FellhauerEtAl2007}
\begin{equation}
\Phi(x,y,z) = \Phi_h(r) + \Phi_d(R,z) + \Phi_b(r)
\end{equation}
where
\begin{equation}
r = \sqrt{x^2 + y^2 + z^2} \quad \quad R = \sqrt{x^2 + y^2}
\end{equation}
and $x$, $y$ and $z$ are galactic coordinates with units in kpc. In this coordinate system the Sun is located at (-8,0,0) at which a particle with velocity directed in the positive $\mathbf{y}$ direction is moving in the direction of Galactic rotation. It follows that $\mathbf{z}$ points in the direction of the northern galactic pole. 
The gravitational potentials for the galactic halo $\Phi_h$, disc $\Phi_d$ and bulge $\Phi_b$ are
\begin{eqnarray}
\Phi_h(r)   &=& \frac{1}{2} v_o^2 \ln \left( 1 + \frac{r^2}{a^2} \right) \\
\Phi_d(R,z) &=& \frac{- G M_d}{ \sqrt{ R^2 + \left( b + \sqrt{z^2 + c^2} \right)^2 } } \\
\Phi_b(r)   &=& \frac{- G M_b}{ r + d}
\end{eqnarray}
where $a$ = 12.0 kpc, $b$ = 6.5 kpc, $c$ = 0.26 kpc, $d$ = 0.7 kpc, $v_o$ = 181 km/s and $G$ is the gravitational constant. The masses of the galactic disc and bulge are $M_d = 10^{11}$ M$_\odot$ and $M_b = 3.4 \times 10^{10}$ M$_\odot$ respectively. The equations of motion for a cluster moving in this potential are given by
\begin{equation}
\mathbf{\ddot{r}} = -\nabla \left(\Phi_h + \Phi_d + \Phi_b \right). 
\label{gc_GGCS_EOM}
\end{equation}
For globular clusters with velocity data given in the literature, \reqOne{gc_GGCS_EOM} is integrated back through time for approximately 10 cluster-galaxy orbits, using the Bulirsch-Stoer numerical integration method \citep{NR86}. Then $r_{min}$ and $r_{max}$ are measured and used to approximate the effective Keplerian orbital eccentricity ($e$) and perigalacticon ($R_p$).

Ideally the orbit for each GC can be determined by integrating the observed velocities once using this method. However the uncertainties in these velocities are quite large, especially the velocity components that strongly depend on proper motion observations. To accurately describe the range of orbits consistent with this range of velocities, 1000 realisations are run for each cluster. Each realisation consists of a different 3-component velocity chosen from the error bars of the observed radial velocity and tangential velocity components derived by proper motions in right ascension and declination. The errors in these radial velocity and proper motion observations are assumed to be normally distributed with standard deviation taken from the literature error bars. The 3-component velocity is then converted into galactic coordinates and integrated back through time to determine the approximate orbit ($R_p$, $e$). 

\begin{figure}
\begin{centering}\resizebox{\figsmall}{!}{\includegraphics[angle=270]{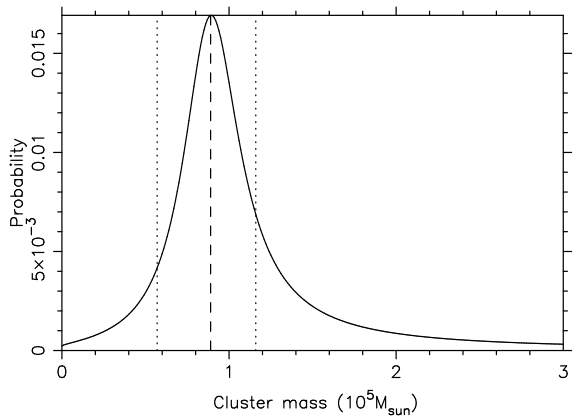}} \par\end{centering}
\caption
{Probability function for the cluster mass of NGC 6809 determined by fitting the mass to the observed velocity dispersion profile in the core of the cluster (see in-text for details).}
\label{fig:GCMassProb} 
\end{figure}

The initial conditions and resulting orbital parameters from these cluster-galaxy orbit simulations are summarised in Table~\ref{TblObs}. Column 2 lists the cluster mass estimated from the different values found in the literature ($M_{C,lit}$). The fitted cluster masses in column 3 of Table~\ref{TblObs} are determined using a least squares fit to the observed velocity dispersion data and applying \reqTwo{SigmaR}{MCSigma} to the central value. The fitted masses are determined from the mass with the highest probability of fitting the velocity dispersion data (dashed line), while the uncertainties are calculated as where there is a 34.1\% probability for the cluster mass to be in the positive or negative direction from the best fit (dotted lines). If the cluster masses were normally distributed then this would be equivalent to one standard deviation in each direction. An example probability distribution normalised to sum to unity for the total cluster mass of NGC 6809 is shown in Figure~\ref{fig:GCMassProb}. The distribution of cluster masses is found to be best described as a skewed normal distribution. Note that by estimating the cluster mass and associated uncertainties in this way there is no need to add an additional measurement error due to the uncertainty in the cluster distance.

\begin{table*}
 \caption{Globular cluster orbital parameters and uncertainties for 15 Milky Way globular clusters.}
 \label{TblObs}
\begin{centering}
 \begin{tabular}{lcccccccc}
\hline 
\hline 
Cluster & $M_{C,lit}$ & $M_{C,fit}$ & $R_{p}$ & $R_{a}$ & $e$ & $\Phi$ & $f(t)_{a<a_0}$ & References \\
 & ($\times 10^5 \msun$) & ($\times 10^5\msun$) & (kpc)  & (kpc)  &  &  &   &  \\
\hline 
 \vspace{\tablegap}
NGC   104 & $    11.3\pm     2.7$ &     8.98$^{+    2.15}_{-    2.27}$ &      4.1$^{+     0.2}_{-     0.2}$ &      7.9$^{+     0.2}_{-     0.1}$ &     0.32$^{+    0.03}_{-    0.03}$ &    198.0$^{+     1.0}_{-     1.0}$ &     0.00$^{+    0.00}_{-    0.00}$ &    H,7,9,11,12,14,16-18  \\  \vspace{\tablegap}
NGC   288 & $     0.8\pm     0.3$ &     0.87$^{+    0.19}_{-    0.21}$ &      3.0$^{+     1.5}_{-     1.2}$ &     11.9$^{+     0.7}_{-     0.7}$ &     0.60$^{+    0.14}_{-    0.12}$ &    176.8$^{+     1.2}_{-     2.7}$ &     0.38$^{+    0.06}_{-    0.04}$ &    E,2,3,11,12,14,16-18  \\  \vspace{\tablegap}
NGC  1851 & $     4.1\pm     1.4$ &     3.74$^{+    0.06}_{-    0.08}$ &      1.1$^{+     0.7}_{-     0.4}$ &     27.0$^{+     5.0}_{-     3.5}$ &     0.92$^{+    0.02}_{-    0.05}$ &    201.2$^{+    17.5}_{-    10.9}$ &     0.68$^{+    0.10}_{-    0.13}$ &    I,3,12,14,16,18,19    \\  \vspace{\tablegap}
NGC  1904 & $     2.1\pm     1.0$ &     1.37$^{+    0.32}_{-    0.34}$ &      2.3$^{+     1.6}_{-     1.0}$ &     20.6$^{+     1.6}_{-     1.5}$ &     0.80$^{+    0.08}_{-    0.10}$ &    201.4$^{+    25.5}_{-    12.0}$ &     0.56$^{+    0.20}_{-    0.23}$ &    I,3,12,14,15,19       \\  \vspace{\tablegap}
NGC  5024 & $     6.1\pm     1.5$ &     5.00$^{+    1.16}_{-    1.22}$ &     15.3$^{+     3.1}_{-     5.1}$ &     25.5$^{+    14.5}_{-     5.1}$ &     0.32$^{+    0.14}_{-    0.11}$ &     25.4$^{+    36.7}_{-    13.1}$ &     0.93$^{+    0.06}_{-    0.13}$ &    H,4,11,16,18          \\  \vspace{\tablegap}
NGC  5139 & $    29.1\pm    12.7$ &    34.23$^{+    6.03}_{-    6.36}$ &      1.0$^{+     0.2}_{-     0.2}$ &      6.3$^{+     0.1}_{-     0.0}$ &     0.72$^{+    0.05}_{-    0.05}$ &    177.9$^{+     0.1}_{-     0.2}$ &     0.00$^{+    0.00}_{-    0.00}$ &    B,G,6,11,12,14,16-19  \\  \vspace{\tablegap}
NGC  6121 & $     1.1\pm     0.5$ &     1.46$^{+    0.42}_{-    0.43}$ &      0.6$^{+     0.2}_{-     0.2}$ &      5.5$^{+     0.2}_{-     0.2}$ &     0.81$^{+    0.05}_{-    0.07}$ &    179.2$^{+     0.5}_{-     0.8}$ &     0.00$^{+    0.05}_{-    0.00}$ &    H,6,8,11,14,16-18     \\  \vspace{\tablegap}
NGC  6171 & $     1.1\pm     0.4$ &     0.98$^{+    0.15}_{-    0.16}$ &      2.0$^{+     0.7}_{-     0.6}$ &      3.5$^{+     0.2}_{-     0.2}$ &     0.27$^{+    0.17}_{-    0.16}$ &    177.6$^{+     1.9}_{-     2.4}$ &     0.00$^{+    0.00}_{-    0.00}$ &    C,1,11,14,16,18,19    \\  \vspace{\tablegap}
NGC  6218 & $     1.3\pm     0.5$ &     1.04$^{+    0.12}_{-    0.12}$ &      1.0$^{+     0.4}_{-     0.3}$ &      5.3$^{+     0.3}_{-     0.3}$ &     0.68$^{+    0.09}_{-    0.10}$ &    174.8$^{+     2.0}_{-     2.3}$ &     0.00$^{+    0.00}_{-    0.00}$ &    H,4,11,14,16-18       \\  \vspace{\tablegap}
NGC  6341 & $     2.1\pm     0.6$ &     1.85$^{+    0.30}_{-    0.31}$ &      1.6$^{+     0.5}_{-     0.2}$ &      9.8$^{+     0.6}_{-     0.5}$ &     0.72$^{+    0.02}_{-    0.06}$ &    183.8$^{+     3.8}_{-     3.3}$ &     0.20$^{+    0.25}_{-    0.20}$ &    D,1,5,11,14,16,17,19  \\  \vspace{\tablegap}
NGC  6656 & $     4.4\pm     1.2$ &     3.18$^{+    1.00}_{-    1.07}$ &      3.1$^{+     0.3}_{-     0.3}$ &      8.3$^{+     0.8}_{-     0.6}$ &     0.45$^{+    0.04}_{-    0.02}$ &    223.6$^{+    11.1}_{-     8.3}$ &     0.00$^{+    0.00}_{-    0.00}$ &    H,1,11,14,16-18       \\  \vspace{\tablegap}
NGC  6752 & $     2.2\pm     0.7$ &     1.76$^{+    0.52}_{-    0.55}$ &      4.1$^{+     0.4}_{-     0.3}$ &      5.3$^{+     0.3}_{-     0.2}$ &     0.13$^{+    0.04}_{-    0.03}$ &    186.3$^{+     1.6}_{-     1.5}$ &     0.00$^{+    0.00}_{-    0.00}$ &    H,3,16-18             \\  \vspace{\tablegap}
NGC  6809 & $     2.0\pm     0.5$ &     0.89$^{+    0.27}_{-    0.32}$ &      1.8$^{+     0.3}_{-     0.3}$ &      5.5$^{+     0.4}_{-     0.3}$ &     0.52$^{+    0.08}_{-    0.09}$ &    164.5$^{+     2.7}_{-     1.9}$ &     0.28$^{+    0.02}_{-    0.07}$ &    H,6,12,14,16-18       \\  \vspace{\tablegap}
NGC  7078 & $     6.8\pm     2.7$ &     3.98$^{+    0.91}_{-    0.99}$ &      5.7$^{+     2.0}_{-     2.1}$ &     20.5$^{+     9.2}_{-     4.9}$ &     0.60$^{+    0.12}_{-    0.12}$ &     76.8$^{+    33.4}_{-    22.4}$ &     0.74$^{+    0.11}_{-    0.15}$ &    A,4,11,13,16-19       \\  \vspace{\tablegap}
NGC  7099 & $     1.5\pm     0.6$ &     0.84$^{+    0.13}_{-    0.15}$ &      3.5$^{+     1.2}_{-     0.9}$ &      7.1$^{+     0.4}_{-     0.3}$ &     0.34$^{+    0.11}_{-    0.11}$ &    179.3$^{+     4.5}_{-     3.9}$ &     0.25$^{+    0.07}_{-    0.08}$ &    F,H,6,14,16-19        \\
\hline
 \end{tabular}
\end{centering}

\medskip
Observable parameters, including errors, for the cluster masses and orbits around the galaxy are given for the 15 GCs in this sample. The cluster mass range from the literature and obtained by fitting to the velocity dispersion in the Newtonian regime (see in text) are given in columns 2 and 3 respectively. The galactic orbit determined from a sample of 1000 initial conditions per cluster (see in text for details) are characterised by the perigalacticon ($R_p$), apogalacticon ($R_a$), eccentricity ($e$), phase ($\Phi$) and the fraction of time the GC is subject to acceleration less than the MOND acceleration. The phase is defined such that 0 is at perigalacticon and positive (negative) angle near this is moving towards (away from) perigalacticon. All GC positions are from \cite{Harris1996} (2010 edition) with additional data sources from the literature given in the final column. The references for the velocity dispersion are; A = \citet{DSCLBMS1998}, B = \citet{ScarpaWCen03}, C = \citet{SMG2004}, D = \citet{DCLSBM2007}, E = \citet{SMGC2007}, F = \citet{ScarpaEtAl2007}, G = \citet{SF2010}, H = \citet{LaneEtAl2011}, I = \citet{SMCF2011}. Proper motions references are; 1 = \cite{CH1993}, 2 = \cite{Guo1995}, 3 = \cite{DinescuI}, 4 = \cite{Odenkirchen1997}, 5 = \cite{Geffert1998}, 6 = \cite{DinescuII}, 7 = \cite{AndersonKing2003}, 8 = \cite{BedinEtAl2003}, 9 = \cite{FreireEtAl2003}, 10 = \cite{KaliraiEtAl2004}. References for the cluster masses are; 11 = \cite{Meziane96}, 12 = \cite{LeonEtAl2000}, 13 = \cite{MHB2004}, 14 = \cite{McLvdM2005} (calculated from their central velocity dispersion observational values along with \reqTwoNP{SigmaR}{MCSigma}), 15 = \cite{AMP2006}, 16 = \cite{VPC2009}, 17 = \cite{MK2010}, 18 = \cite{BoylesEtAl2011}, and 19 = \cite{SMCF2011}.
\end{table*}

The masses determined using this method are consistent within error bars to the cluster masses from the literature, which are often inconsistent between studies. The literature cluster masses are an average of the individual mass values given in the papers listed in the reference column of Table~\ref{TblObs}. A fitted mass is used for consistency between clusters and to avoid the observational issues in determining cluster masses discussed further in Section~\ref{obsuncert}. Since the cluster masses also have large uncertainty a similar approach of 1000 realisations drawn from the skewed normal distribution is adopted, making a total of $10^6$ realisations per GC.

Using the 1000 orbital and 1000 cluster mass realisations, Table~\ref{TblObs} lists the median and standard deviation values associated with $34.1\%$ increases in the positive or negative direction. Columns 4-9 list the perigalacticon and apogalacticon distances ($R_p$ and $R_a$), the orbital eccentricity ($e$) orbital phase ($\Phi$), the fraction of time where the acceleration due to the galactic potential is less than the MOND acceleration ($f(t)_{a<a_0}$), and finally the references for orbital and velocity dispersion observational data.

Comparing the orbital parameters $R_p$ and $e$ for individual clusters in Table~\ref{TblObs} to those of \cite{AMP2006} we find our eccentricity value is more radial for NGC 104, NGC 1851, NGC 5139, NGC 6218 and NGC 7078; less radial for NGC 5024 and NGC 6341; the orbital parameters are good fits for NGC 6656 and 7099; and they fit within uncertainties for NGC 288, NGC 1904, NGC 6121, NGC 6171, NGC 6752 and NGC 6809. Of these NGC 1851 and NGC 7078 are known to have very high radial velocity components, which leads to higher eccentricity determinations (see Section~\ref{obsuncert}). Note that \cite{AMP2006} examined three different galactic potential models and found that the orbital parameters vary greatly between galactic potentials. The differences between studies are expected since we are using more recent observed proper motions for each cluster and a different galactic potential, which is most similar to their axisymmetric model. The galactic potential for the Milky Way is an entire field in itself and is out of the scope of the present work.

\subsection{Globular cluster radii}
\label{GCRadii}

The main focus of this work is comparing the radius where any observed deviation in the velocity dispersion profile occurs to the radius predicted by the stability boundary or MOND models.

The observed half-mass radius, flattening radius and predicted radii for each cluster are given in Table~\ref{TblRadii}. The second column gives the half-mass radius ($r_{h}$) corrected for 2D projection using $r_{h} = 1.195 R_{1/2}$ assuming a Plummer density profile. The projected half-mass radii ($R_{1/2}$) are taken from \cite{Harris1996} (2010 edition) along with the distances to each cluster.

\begin{table*}
 \caption{Comparison between observed and predicted radii for all 15 globular clusters with orbital parameters in Table~\ref{TblObs}.}
 \label{TblRadii}
\begin{centering}
 \begin{tabular}{lccccccccc}
\hline 
Cluster & $r_h$ & $r_f^{lit}$ & $r_{f}$ & $r_{m}$ & $r_t$ & $r_{c}$ & $r_{c}^*$ & $T_{rx}(r_c)/10T_o$   \\
        & (pc)  & (pc)   &  (pc)    & (pc)  &  (pc)     & (pc)    &  (pc)  &     \\
\hline 
 \vspace{\tablegap}
NGC   104 &     4.95 &     - &     13.8$^{+     8.7}_{-     5.3}$ &     31.5$^{+     3.7}_{-     4.5}$ &     57.2$^{+     5.3}_{-     5.6}$ &     28.3$^{+     2.8}_{-     2.9}$ &     28.3$^{+     2.8}_{-     2.9}$ &    185.2$^{+    57.0}_{-    49.9}$ \\  \vspace{\tablegap}
NGC   288 &     6.89 &      0.0 &     11.8$^{+     1.2}_{-     7.8}$ &      0.0$^{+     0.0}_{-     0.0}$ &     18.1$^{+     9.5}_{-     7.7}$ &      8.0$^{+     4.8}_{-     3.8}$ &      8.7$^{+     4.1}_{-     0.9}$ &      0.3$^{+     1.0}_{-     0.3}$ \\  \vspace{\tablegap}
NGC  1851 &     2.14 &     12.5 &     16.0$^{+     0.5}_{-     0.5}$ &     20.5$^{+     0.2}_{-     0.2}$ &     10.8$^{+     7.4}_{-     3.7}$ &      3.5$^{+     2.9}_{-     1.3}$ &      4.7$^{+     1.7}_{-     0.4}$ &      0.4$^{+     2.5}_{-     0.3}$ \\  \vspace{\tablegap}
NGC  1904 &     2.91 &     12.0 &     19.2$^{+     1.6}_{-     1.6}$ &     11.9$^{+     1.5}_{-     1.7}$ &     15.8$^{+    11.9}_{-     7.2}$ &      6.1$^{+     5.4}_{-     3.1}$ &      6.2$^{+     5.3}_{-     0.8}$ &      1.2$^{+     8.3}_{-     1.1}$ \\  \vspace{\tablegap}
NGC  5024 &     8.14 &     - &     19.9$^{+    12.0}_{-     3.0}$ &     21.7$^{+     2.9}_{-     3.6}$ &     99.9$^{+     0.0}_{-     0.0}$ &     84.8$^{+    15.1}_{-    29.3}$ &     84.8$^{+    15.1}_{-    29.3}$ &    946.6$^{+  1056.2}_{-   639.8}$ \\  \vspace{\tablegap}
NGC  5139 &     9.03 &     32.0 &     32.6$^{+     3.1}_{-     5.8}$ &     61.8$^{+     5.4}_{-     6.2}$ &     20.5$^{+     5.0}_{-     4.6}$ &      8.4$^{+     2.4}_{-     2.2}$ &      8.4$^{+     2.4}_{-     2.2}$ &     10.6$^{+    14.0}_{-     6.8}$ \\  \vspace{\tablegap}
NGC  6121 &     3.31 &     - &      7.3$^{+     5.0}_{-     1.7}$ &     12.2$^{+     1.8}_{-     2.2}$ &      4.1$^{+     1.6}_{-     1.2}$ &      1.5$^{+     0.7}_{-     0.5}$ &      3.3$^{+     0.2}_{-     0.4}$ &      0.1$^{+     0.2}_{-     0.0}$ \\  \vspace{\tablegap}
NGC  6171 &     3.84 &      8.0 &      7.6$^{+     1.9}_{-     1.8}$ &      9.4$^{+     0.8}_{-     1.1}$ &     13.0$^{+     5.1}_{-     4.2}$ &      6.5$^{+     3.0}_{-     2.4}$ &      6.5$^{+     3.0}_{-     2.4}$ &      6.3$^{+    13.4}_{-     4.8}$ \\  \vspace{\tablegap}
NGC  6218 &     2.95 &     - &      7.3$^{+     1.3}_{-     1.0}$ &     10.2$^{+     0.7}_{-     0.7}$ &      6.6$^{+     2.9}_{-     2.1}$ &      2.7$^{+     1.5}_{-     1.0}$ &      3.4$^{+     0.8}_{-     0.1}$ &      0.5$^{+     1.3}_{-     0.4}$ \\  \vspace{\tablegap}
NGC  6341 &     2.94 &     12.0 &     11.8$^{+     1.9}_{-     2.3}$ &     14.0$^{+     1.2}_{-     1.3}$ &     12.6$^{+     3.9}_{-     1.6}$ &      5.1$^{+     1.9}_{-     0.7}$ &      5.1$^{+     1.9}_{-     0.7}$ &      2.3$^{+     3.7}_{-     0.9}$ \\  \vspace{\tablegap}
NGC  6656 &     3.73 &     - &      6.8$^{+     3.7}_{-     5.8}$ &     18.5$^{+     2.9}_{-     3.7}$ &     29.4$^{+     4.6}_{-     4.8}$ &     13.8$^{+     2.2}_{-     2.4}$ &     13.8$^{+     2.2}_{-     2.4}$ &     54.2$^{+    41.2}_{-    28.5}$ \\  \vspace{\tablegap}
NGC  6752 &     2.65 &     - &      6.8$^{+    17.8}_{-     3.3}$ &     13.7$^{+     2.1}_{-     2.5}$ &     33.7$^{+     4.7}_{-     4.8}$ &     17.5$^{+     2.7}_{-     2.5}$ &     17.5$^{+     2.7}_{-     2.5}$ &    244.4$^{+   168.8}_{-   119.0}$ \\  \vspace{\tablegap}
NGC  6809 &     5.30 &     - &      2.3$^{+    10.3}_{-     2.3}$ &      7.3$^{+     2.0}_{-     7.3}$ &     10.7$^{+     2.5}_{-     2.2}$ &      4.8$^{+     1.4}_{-     1.1}$ &      5.5$^{+     0.8}_{-     0.3}$ &      0.7$^{+     1.2}_{-     0.5}$ \\  \vspace{\tablegap}
NGC  7078 &     3.61 &     20.0 &     24.3$^{+     5.6}_{-     4.3}$ &     20.9$^{+     2.3}_{-     3.0}$ &     57.4$^{+    21.5}_{-    21.9}$ &     25.7$^{+    10.2}_{-    10.6}$ &     25.7$^{+    10.2}_{-    10.6}$ &    139.5$^{+   277.4}_{-   110.1}$ \\  \vspace{\tablegap}
NGC  7099 &     2.90 &     10.0 &     10.5$^{+     0.6}_{-     3.1}$ &      9.0$^{+     0.8}_{-     1.0}$ &     21.9$^{+     8.0}_{-     5.6}$ &     10.7$^{+     4.5}_{-     3.1}$ &     10.7$^{+     4.5}_{-     3.1}$ &     21.5$^{+    38.4}_{-    13.8}$ \\
\hline
 \end{tabular}
\end{centering}

\medskip
For each cluster the observed half mass radius ($r_h$), the approximate radius at which the velocity dispersion flattens by the literature ($r_f^{lit}$) and the fitted value ($r_f$) is given. The predicted radii for the MOND ($r_{m}$) model, tidal radius ($r_t$), chaos radius ($r_c$) and the corrected chaos radius ($r_c^*$). The final column gives the ratio of the relaxation timescale at the chaos radius to the chaotic diffusion timescale which is approximately ten GC-galaxy orbital periods.
\end{table*}

The flattening radius from the literature is given in column 3 of Table~\ref{TblRadii} for each cluster, with values taken from \cite{SMCF2011} for all clusters except NGC 288 which is from \cite{SMGC2007}. These values were based on a by-eye fit to the data, so a more rigorous method to determine the flattening radius is adopted here. We use a simple model for the velocity dispersion which consists of the equilibrium dependence (given by \reqNP{SigmaR}) modified such that $\sigma(r) = \sigma(r_f)$ for $r>r_f$. A best fit is then obtained for $r_f$ for each of the 1000 orbital realisations using a  least squares fit. The resulting median flattening radius and uncertainties are shown in column 4 of Table~\ref{TblRadii}, these are generally in agreement with the literature values, with the exception of NGC 288 and to a lesser extent NGC 1851 and NGC 1904.

Predicted radii based on the GC mass and orbital parameters (given in Table~\ref{TblObs}) are shown in columns 5-8 of Table~\ref{TblRadii} for the MOND acceleration model ($r_{m}$), the tidal radius ($r_t$) and the stability boundary ($r_c$ and $r_c^*$). The tidal radius values presented in this table can be quite different compared to \citet{AMP2006} (and therefore Hernandez et al 2012) due to the differences in the GC orbits (discussed previously).

Recall from Section~\ref{GCRadPred} that the chaos radius alone will not predict the occurrence of escaping stars and therefore a deviation from the velocity dispersion profile. There must be sufficient time for a star on an unstable orbit to escape the cluster compared to the local relaxation timescale. Therefore the radius associated with the stability boundary ($r_c$ given by \reqTwoNP{rtidal}{rtidal_MSC}) needs to be corrected such that the escape timescale ($\approx 10 T_o$) is less than or equal to the relaxation timescale (\reqNP{gc_relaxtime}). The corrected chaos radius, $r_c^*$ (defined in Section~\ref{CorrectedRc}) and is the predicted radius for velocity dispersion flattening by the stability boundary model. The ratio between the relaxation timescale at $r_c$ and ten GC-galaxy orbital periods is shown in the final column of Table~\ref{TblRadii}. When this ratio is less than unity then $r_c^* \sim r_c$ and when this is much greater than unity then $r_c^* \sim r_t$.

\begin{figure}
\begin{centering}\resizebox{\figsmall}{!}{\includegraphics[angle=270]{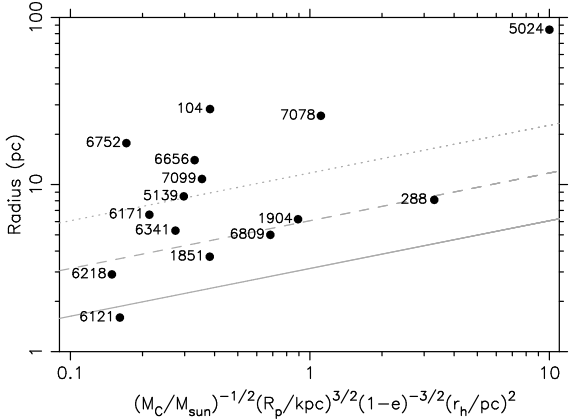}} \par\end{centering}
\caption
{The median uncorrected chaos radius for each cluster ($r_c$ from Table~\ref{TblRadii}) against a combination of observed parameters for each of the 15 GCs. The x-axis is a dimensionless unit which summarises the observational parameters relevant to the relaxation timescale for each GC (see \reqNP{gc_relaxtime}). The curves show the radius where $T_{rx} = 1 T_o$ (solid grey curve), $T_{rx} = 10 T_o$ (dashed) and $T_{rx} = 100 T_o$ (dotted). As an example; if a point is beneath the $T_{rx} = T_{escape} = 10 T_o$ curve then it means that the escape timescale is shorter than the relaxation time.}
\label{fig:GCtimes} 
\end{figure}

The uncorrected chaos radius for each cluster is shown against a combination of observed GC parameters in Figure~\ref{fig:GCtimes}. The units of the x-axis are chosen based on the relaxation timescale given by \reqOne{gc_relaxtime} such that the relaxation timescale is a simple function. The curves in Figure~\ref{fig:GCtimes} show the radii where $T_{rx} = 1 T_o$ (solid grey curve), $T_{rx} = 10 T_o$ (dashed) and $T_{rx} = 100 T_o$ (dotted). Since the escape timescale is approximately 10 GC-galaxy orbital periods, most GCs have the chaotic diffusion timescale is shorter than the relaxation timescale. For NGC 6162, NGC 6218 and NGC 6809 some correction is required so that the chaos radius more closely resembles the tidal radius as seen in Table~\ref{TblRadii} and later in Figures~\ref{VD_All_DIST1}-\ref{VD_All_DIST3}.

\subsection{GC velocity dispersion profiles}

\label{VDprofiles}

This study is limited to clusters for which online data is available for the velocity dispersion as a function of radius. We are further limited to clusters where the orbital parameters can be determined. The galactic globular clusters fitting these requirements are listed in Table~\ref{TblObs} along with the data sources for each cluster.

Three special cases for the observed velocity dispersion data exist in this sample. Firstly, the velocity dispersion data for NGC 7078 are taken from the non-rotating cluster model favoured by \citet{DSCLBMS1998}, rather than the rotating cluster model. Secondly, the velocity dispersion data for NGC 6341 \citep{DCLSBM2007} was not binned in radius in the same way as the other clusters in this sample. For this cluster no data was given on the minimum or maximum distances for each bin, nor for the number of stars used in each bin. This means that no radial error bars are available for the NGC 6341 data points. Finally, not enough velocity dispersion data is currently available for NGC 4590 to build up a reliable cluster mass or equilibrium model fit, hence its absence in our GC subset. The velocity dispersion data from \citet{LaneEtAl2011} for NGC 4590 consistently increases with radius from the centre, as this is non-physical it is likely that there are issues with cluster membership for the stars used in the sample.

The bottom panels in Figures~\ref{VD_All_DIST1}-\ref{VD_All_DIST3} show the velocity dispersion profiles for all clusters listed in Table~\ref{TblObs}. The equilibrium velocity dispersion profile given by \reqOne{SigmaR} is plotted in the bottom panels for the highest probability cluster mass, given in column 3 of Table~\ref{TblObs}. Uncertainties in the orbital elements ($R_p$ and $e$) and cluster mass ($M_C$) are treated using the $10^6$ realisations per GC as discussed above. For each set of orbital elements and cluster mass the predicted radii for each model are determined and the resulting cumulative distributions are shown in the top panels of each figure. The MOND radius $r_{m}$ as calculated for each cluster from \reqOne{ra0} is shown in the top panels as a dashed red line. The relaxation timescale corrected transition from stable to unstable orbits $r_{c}^*$ (Section~\ref{CorrectedRc}) is shown as a solid black line, the tidal radius using \reqOne{rtidal_King} is shown as a dotted green line and the best fit flattening radius is shown as a dot-dashed blue line.

\begin{figure*}
\begin{centering}$\begin{array}{cccc}
\multicolumn{1}{l}{\mbox{(a) {\bf NGC 104}; $M_C$ = $8.89 \times 10^{5} \msun$}} & 
\multicolumn{1}{l}{\mbox{(b) {\bf NGC 288}; $M_C$ = $0.87 \times 10^{5} \msun$}}\\[-0.1cm]
\includegraphics[width=\figsizeFour,angle=270]{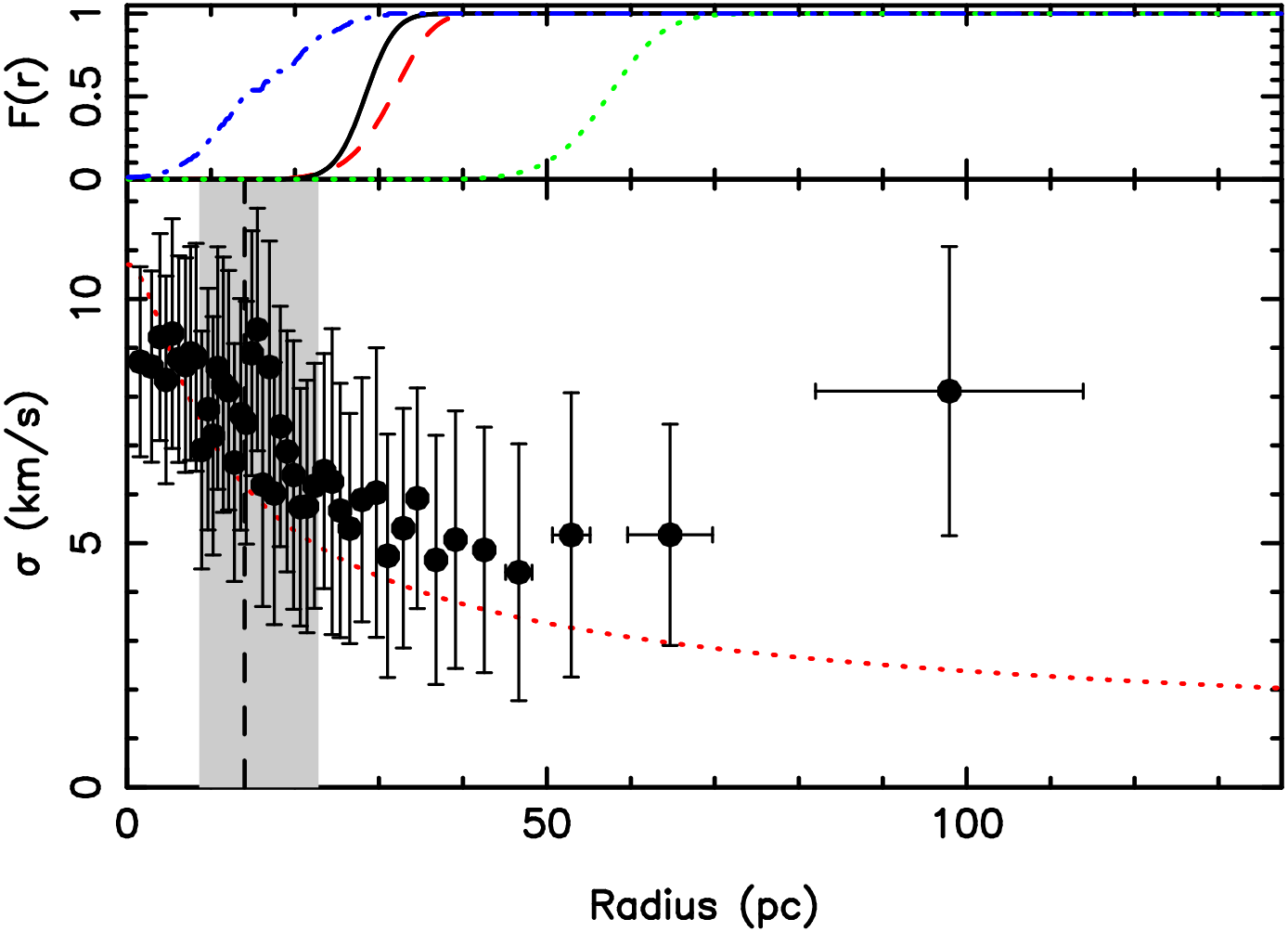} & 
\includegraphics[width=\figsizeFour,angle=270]{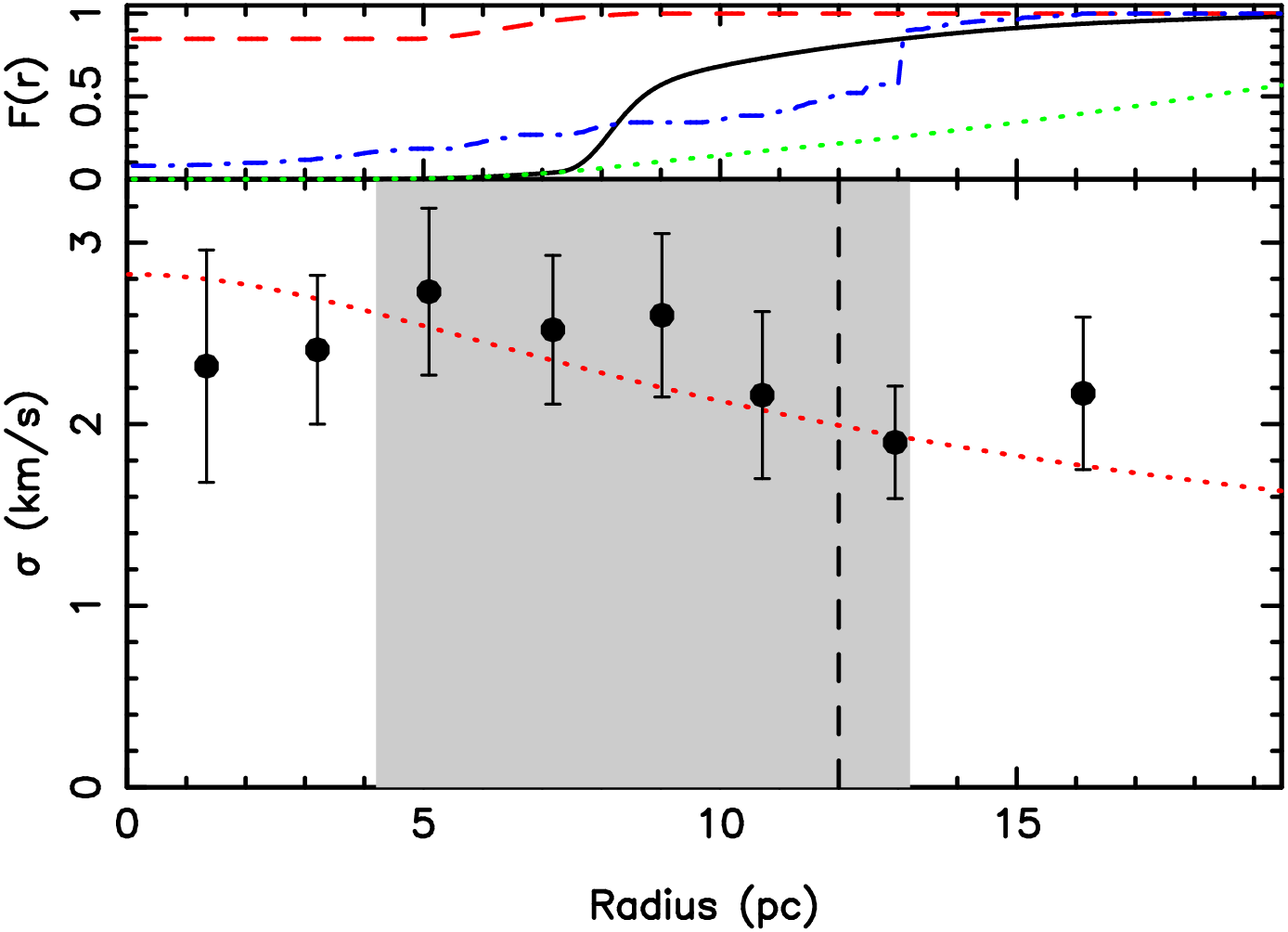}\\
\multicolumn{1}{c}{\mbox{}} & \multicolumn{1}{c}{\mbox{}}\\
\multicolumn{1}{l}{\mbox{(c) {\bf NGC 1851}; $M_C$ = $3.74 \times 10^{5} \msun$}} & 
\multicolumn{1}{l}{\mbox{(d) {\bf NGC 1904}; $M_C$ = $1.37 \times 10^{5} \msun$}}\\[-0.1cm]
\includegraphics[width=\figsizeFour,angle=270]{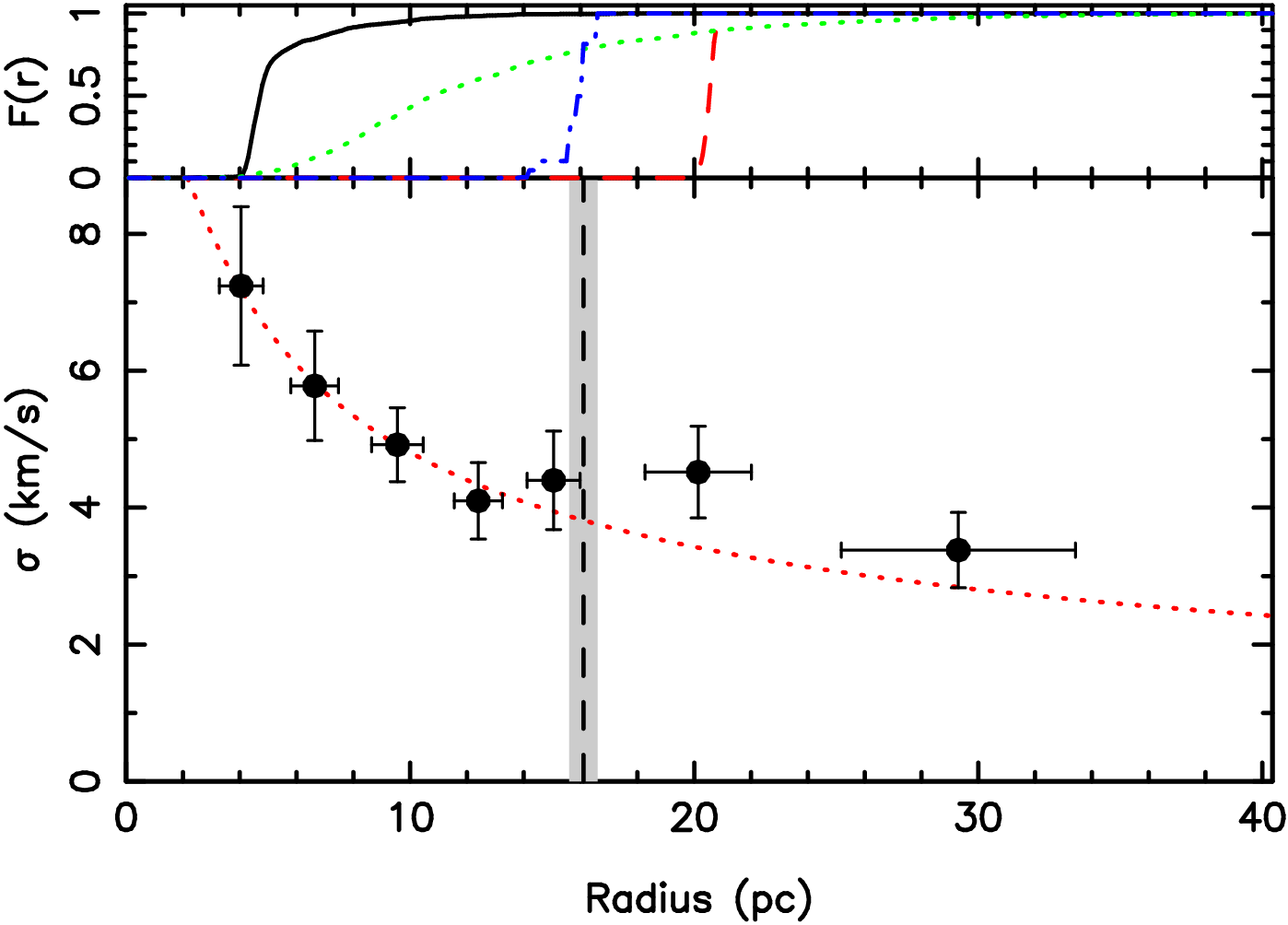} & 
\includegraphics[width=\figsizeFour,angle=270]{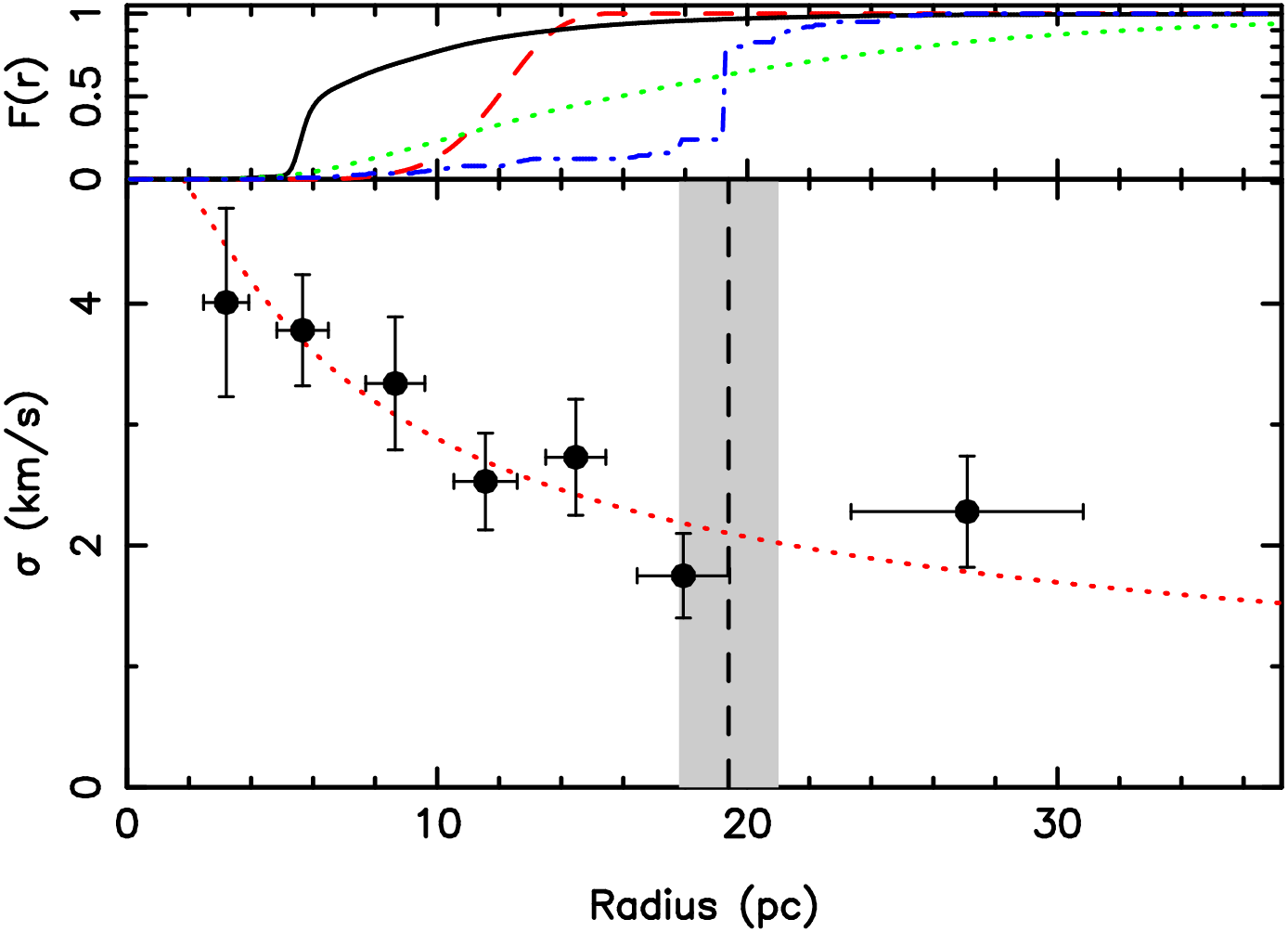}\\
\multicolumn{1}{c}{\mbox{}} & \multicolumn{1}{c}{\mbox{}}\\
\multicolumn{1}{l}{\mbox{(e) {\bf NGC 5024}; $M_C$ = $5.00 \times 10^{5} \msun$}} & 
\multicolumn{1}{l}{\mbox{(f) {\bf NGC 5139}; $M_C$ = $34.23 \times 10^{5} \msun$}}\\[-0.1cm]
\includegraphics[width=\figsizeFour,angle=270]{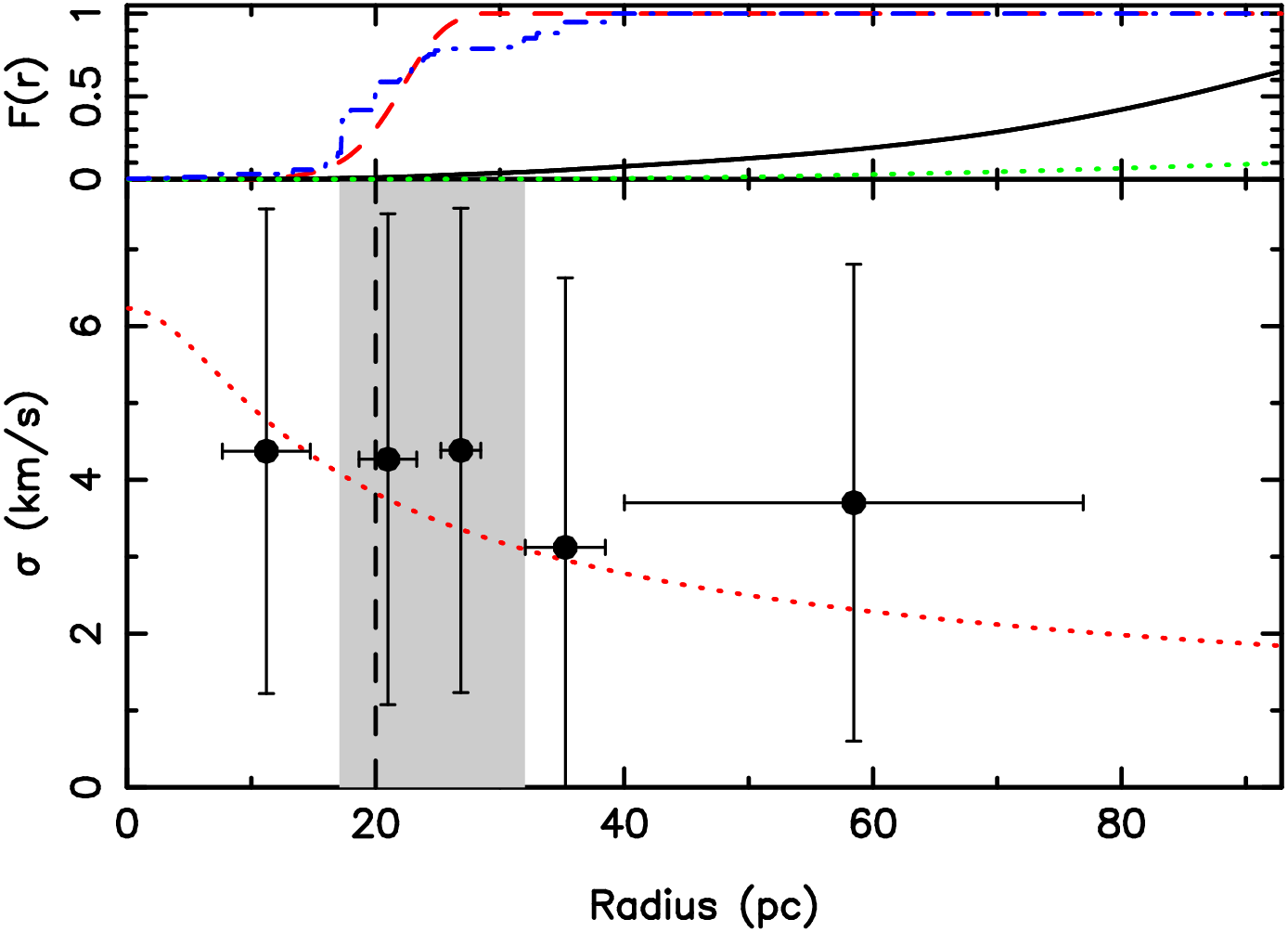} & 
\includegraphics[width=\figsizeFour,angle=270]{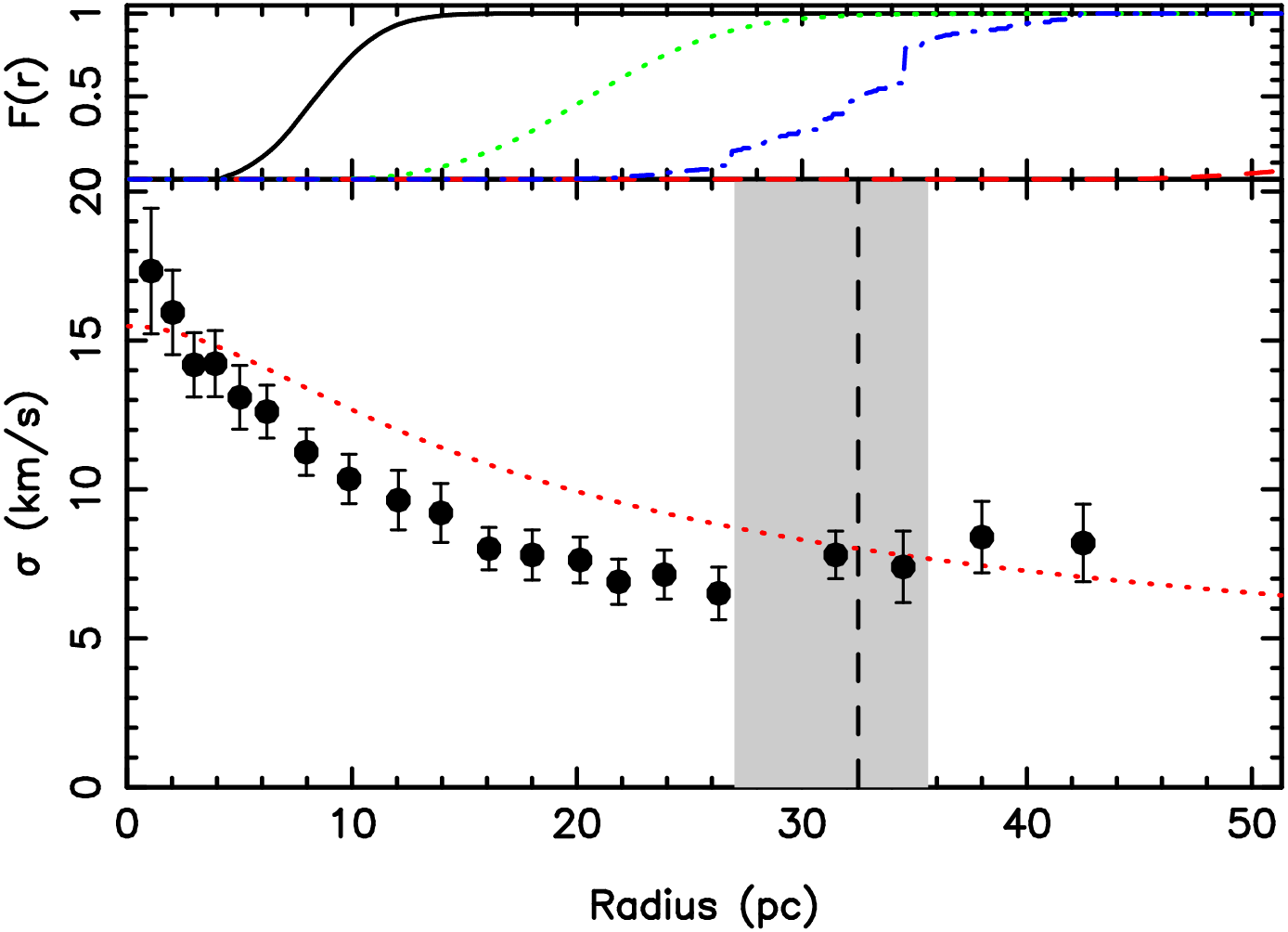}
\end{array}$ \par\end{centering}
\caption{The velocity dispersion profile for all globular clusters in this study using the orbital parameters and total mass distributions given in Table~\ref{TblObs}. The median cluster masses ($M_C$) are shown beside the cluster name. Velocity dispersion observations are taken from the literature, with references in Table~\ref{TblObs}, and are presented with error bars where available. The vertical lines show the best fit to the flattening of the velocity dispersion with errors denoted by the shaded region (as listed in Table~\ref{TblRadii}) and the red dotted curve indicates the theoretical velocity dispersion profile for the quoted cluster mass. The cumulative distributions above each velocity dispersion profile show the chaos radius (solid black curve), MOND radius (red dashed curve), tidal radius (green dotted curve) and best fit flattening radius (dot-dashed blue curve). The larger variation in the chaos and tidal radii for each cluster is due to the uncertainty in the orbital eccentricity.}
\label{VD_All_DIST1} 
\end{figure*}

\begin{figure*}
\begin{centering}$\begin{array}{cccc}
\multicolumn{1}{l}{\mbox{(a) {\bf NGC 6121}; $M_C$ = $1.46 \times 10^{5} \msun$}} & 
\multicolumn{1}{l}{\mbox{(b) {\bf NGC 6171}; $M_C$ = $0.98 \times 10^{5} \msun$}}\\[-0.1cm]
\includegraphics[width=\figsizeFour,angle=270]{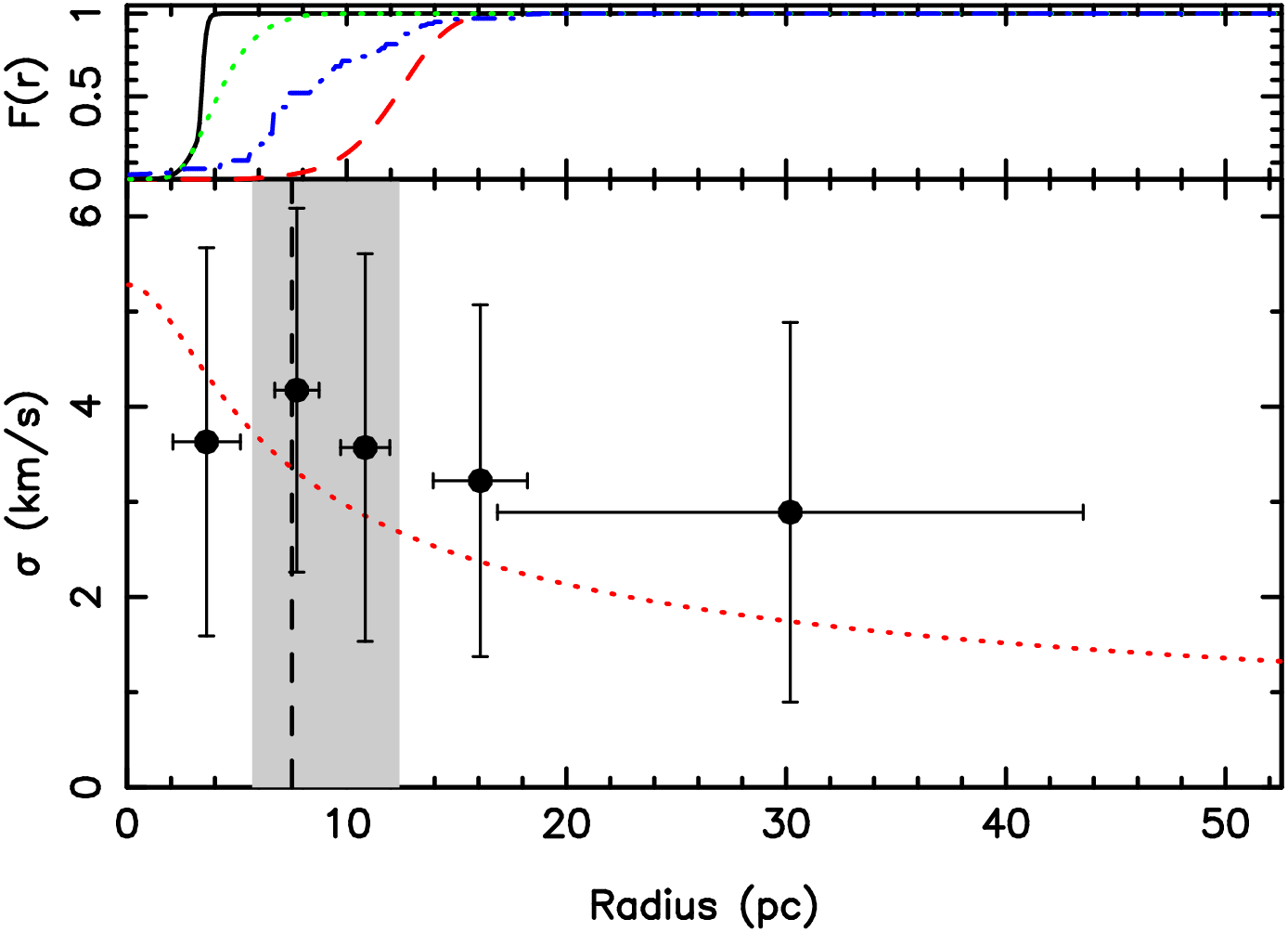} & 
\includegraphics[width=\figsizeFour,angle=270]{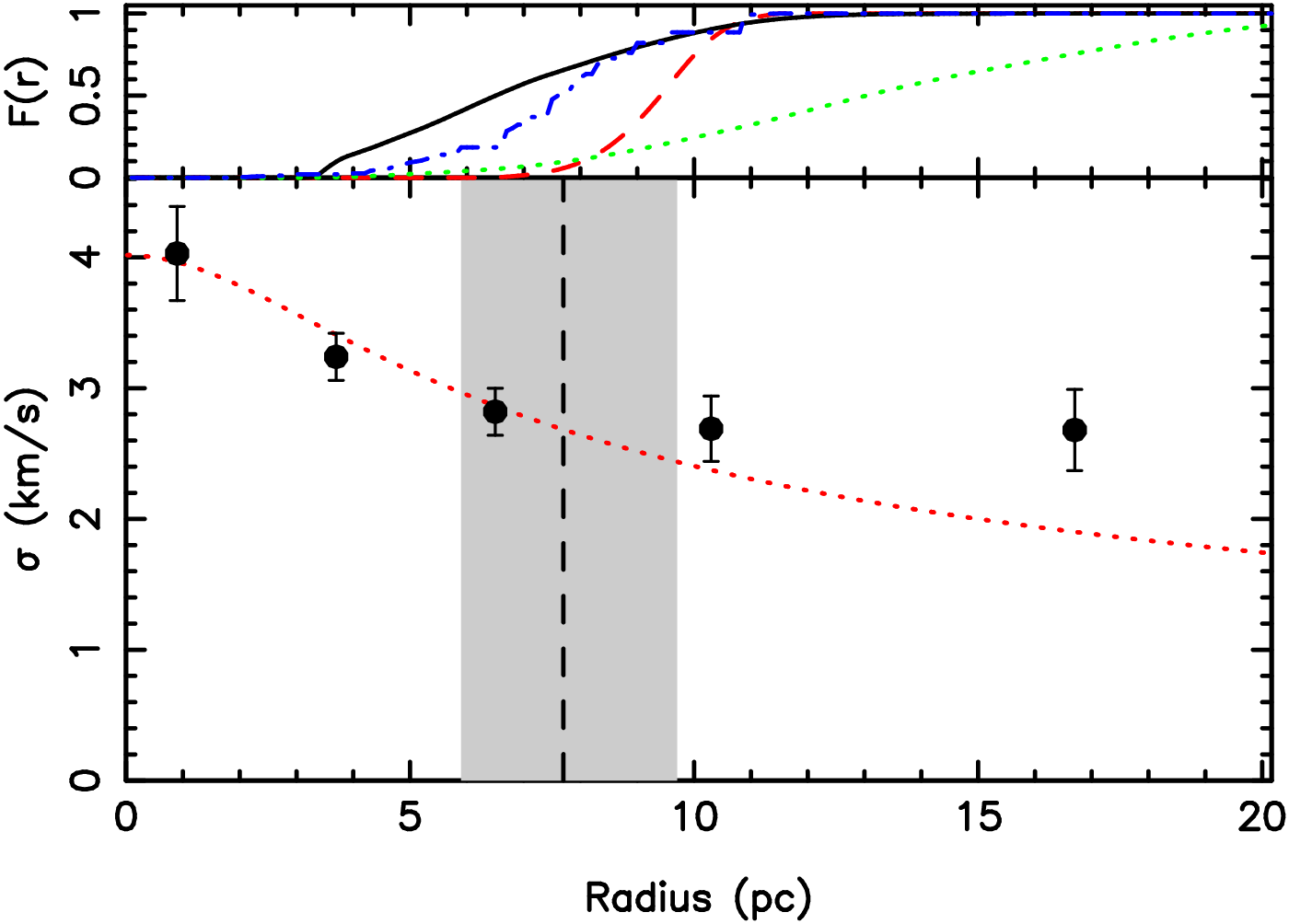}\\
\multicolumn{1}{c}{\mbox{}} & \multicolumn{1}{c}{\mbox{}}\\
\multicolumn{1}{l}{\mbox{(c) {\bf NGC 6218}; $M_C$ = $1.04 \times 10^{5} \msun$}} & 
\multicolumn{1}{l}{\mbox{(d) {\bf NGC 6341}; $M_C$ = $1.85 \times 10^{5} \msun$}}\\[-0.1cm]
\includegraphics[width=\figsizeFour,angle=270]{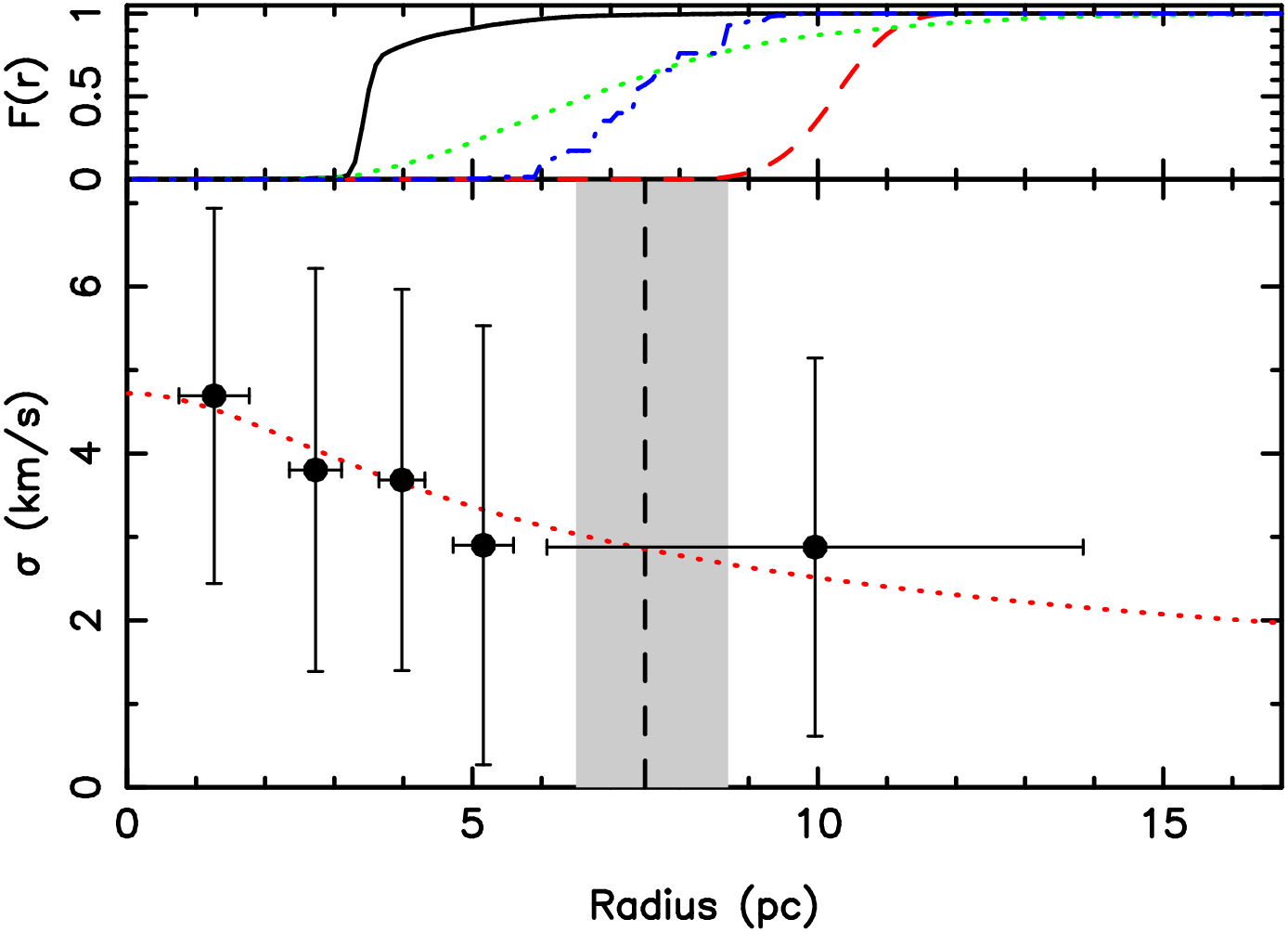} & 
\includegraphics[width=\figsizeFour,angle=270]{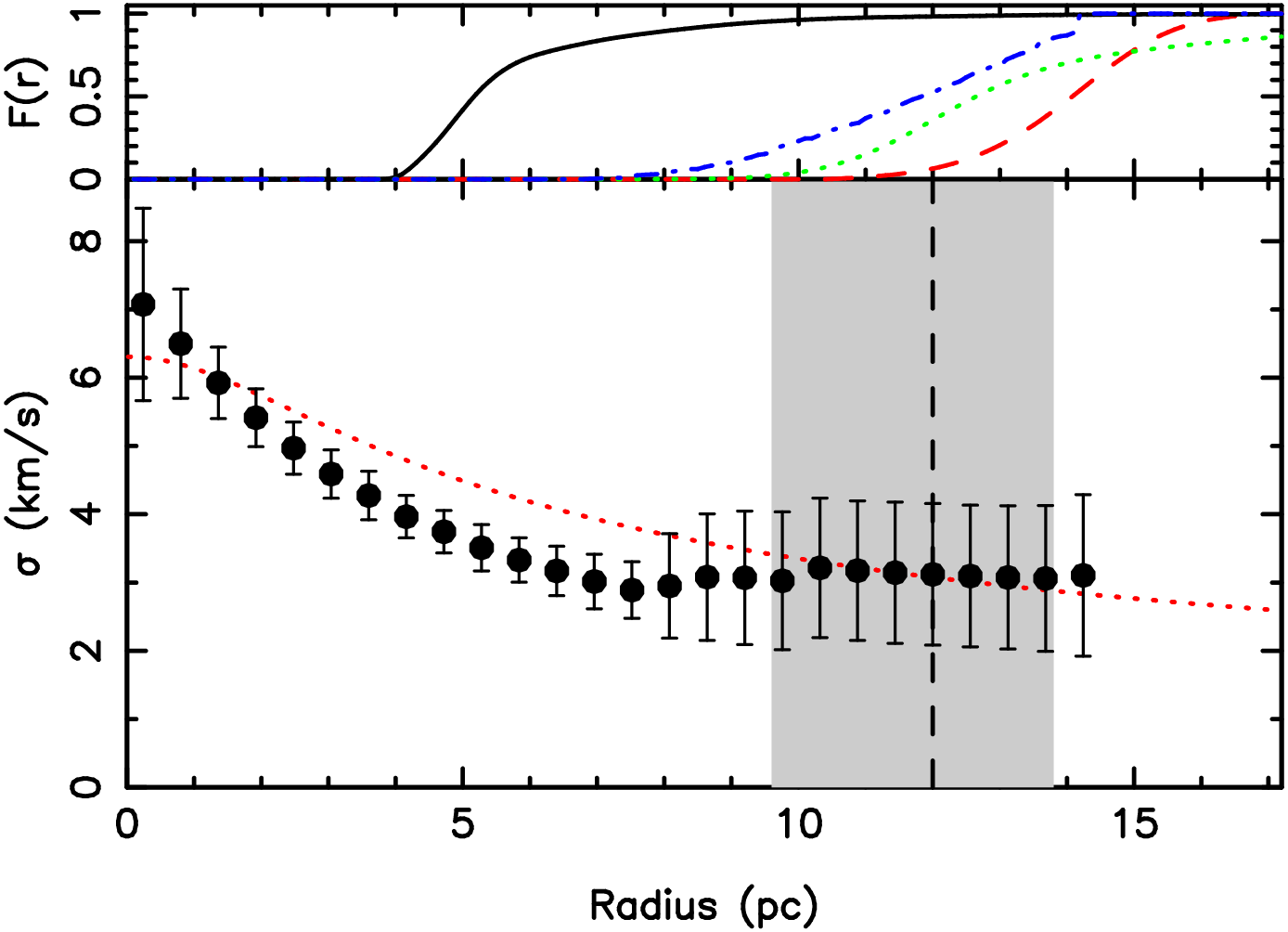}\\
\multicolumn{1}{c}{\mbox{}} & \multicolumn{1}{c}{\mbox{}}\\
\multicolumn{1}{l}{\mbox{(e) {\bf NGC 6656}; $M_C$ = $3.18 \times 10^{5} \msun$}} & 
\multicolumn{1}{l}{\mbox{(f) {\bf NGC 6752}; $M_C$ = $1.76 \times 10^{5} \msun$}}\\[-0.1cm]
\includegraphics[width=\figsizeFour,angle=270]{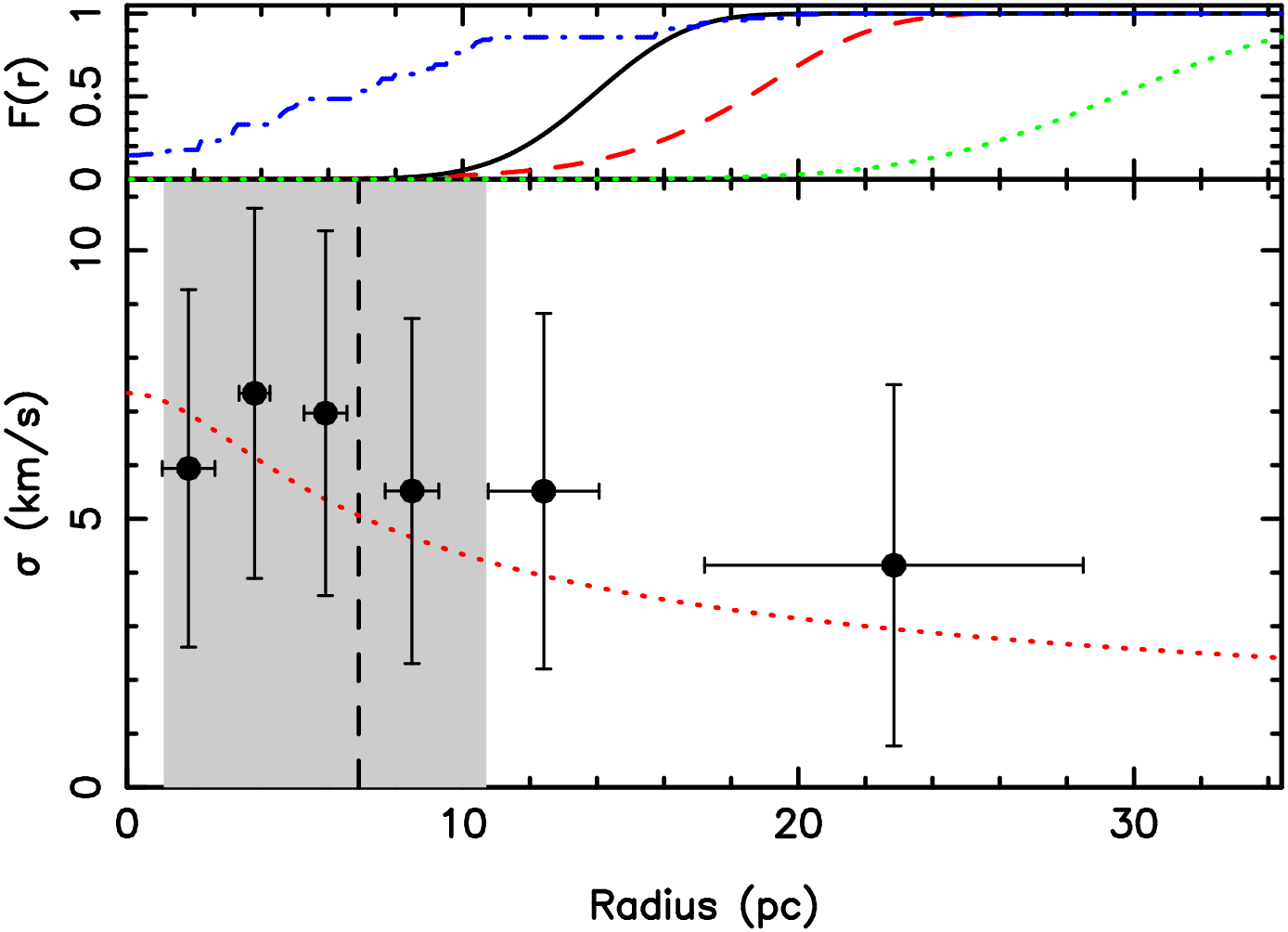} & 
\includegraphics[width=\figsizeFour,angle=270]{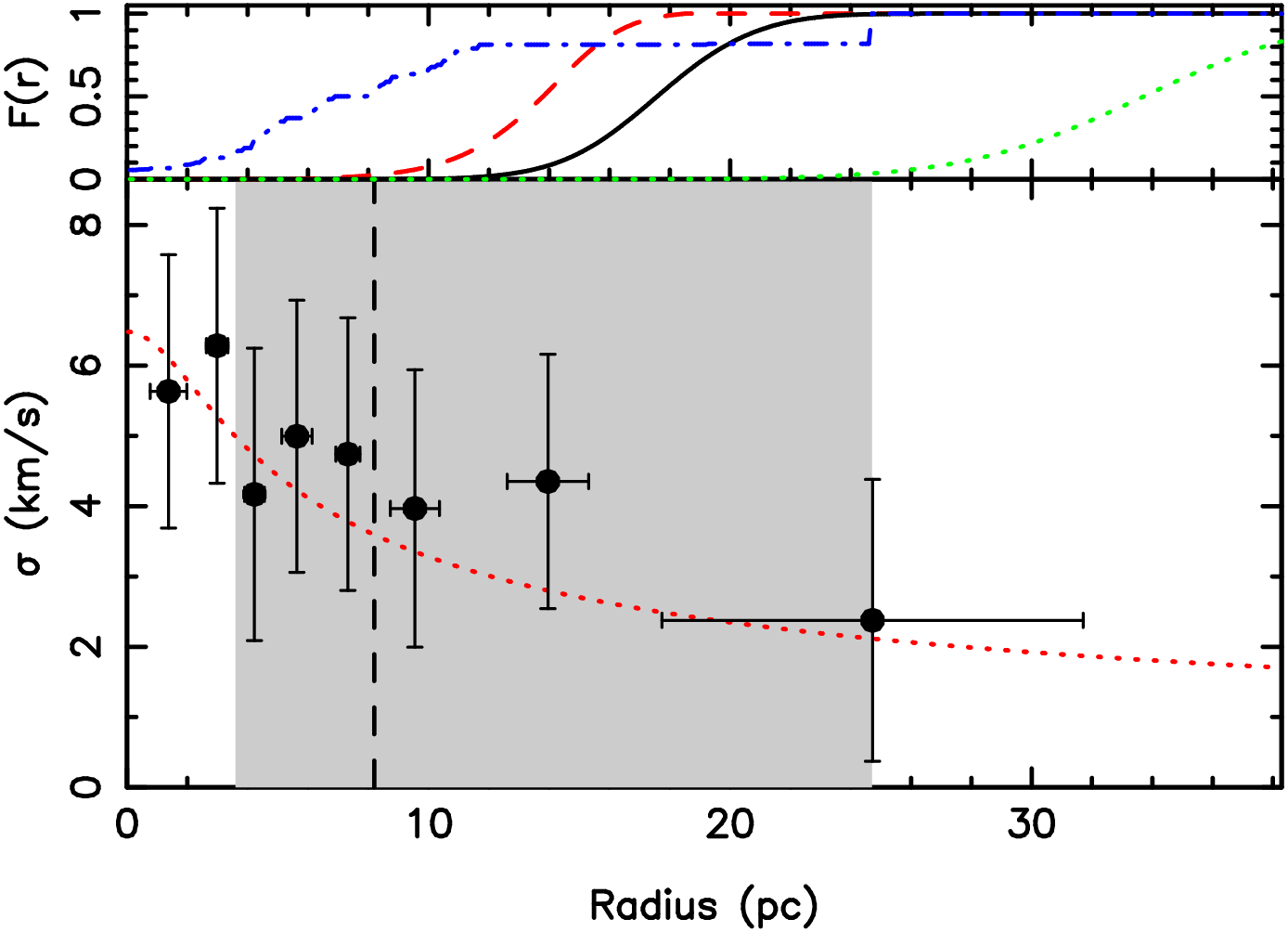}
\end{array}$ \par\end{centering}
\caption{Continuation of Figure~\ref{VD_All_DIST1}.}
\label{VD_All_DIST2} 
\end{figure*}

\begin{figure}
\begin{centering}$\begin{array}{cc}
\multicolumn{1}{l}{\mbox{(a) {\bf NGC 6809}; $M_C$ = $0.89 \times 10^{5} \msun$}} \\[-0.1cm]
\includegraphics[width=\figsizeFour,angle=270]{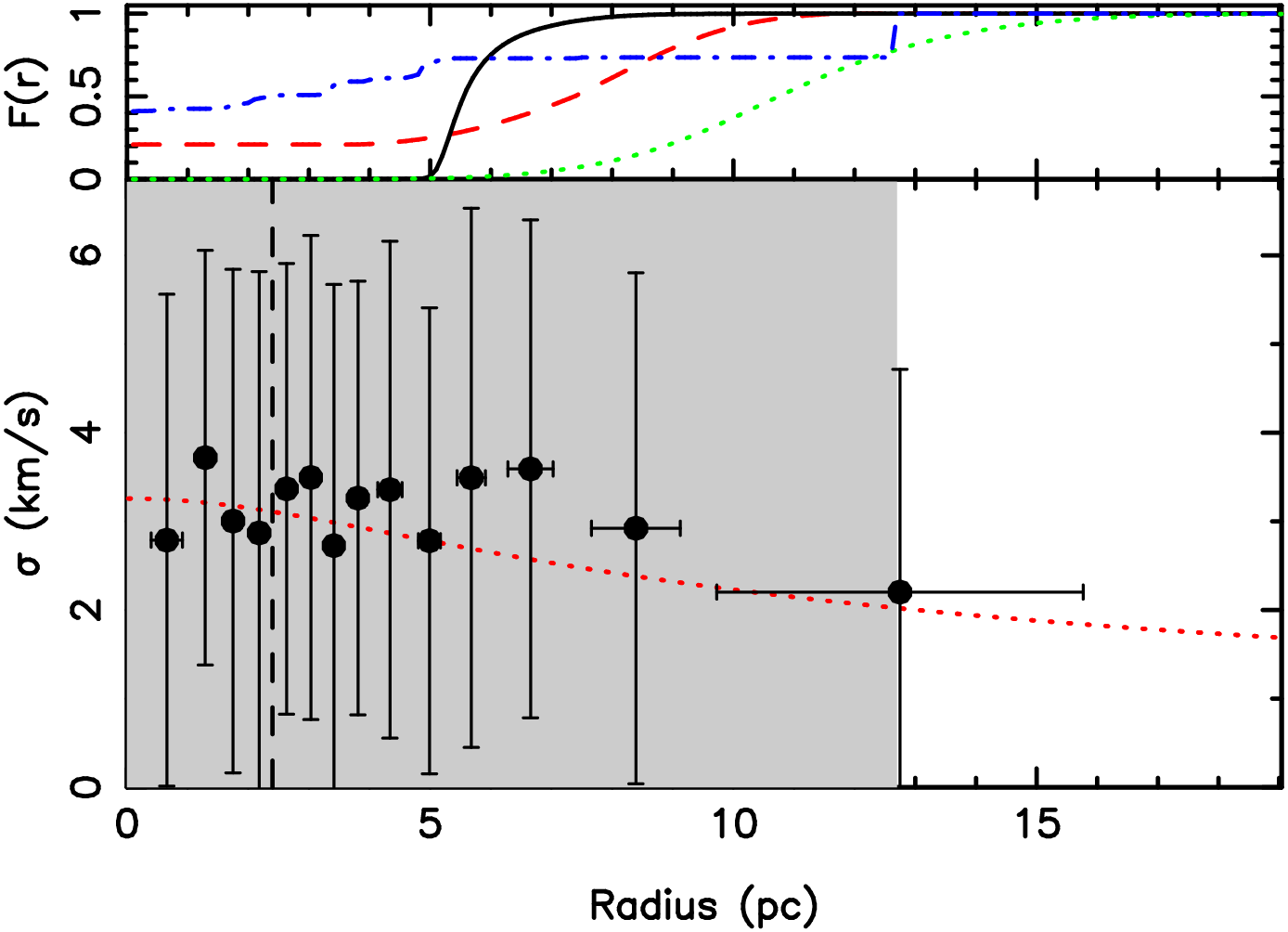} \\
\multicolumn{1}{c}{\mbox{}}  \\
\multicolumn{1}{l}{\mbox{(c) {\bf NGC 7078}; $M_C$ = $3.98 \times 10^{5} \msun$}} \\[-0.1cm]
\includegraphics[width=\figsizeFour,angle=270]{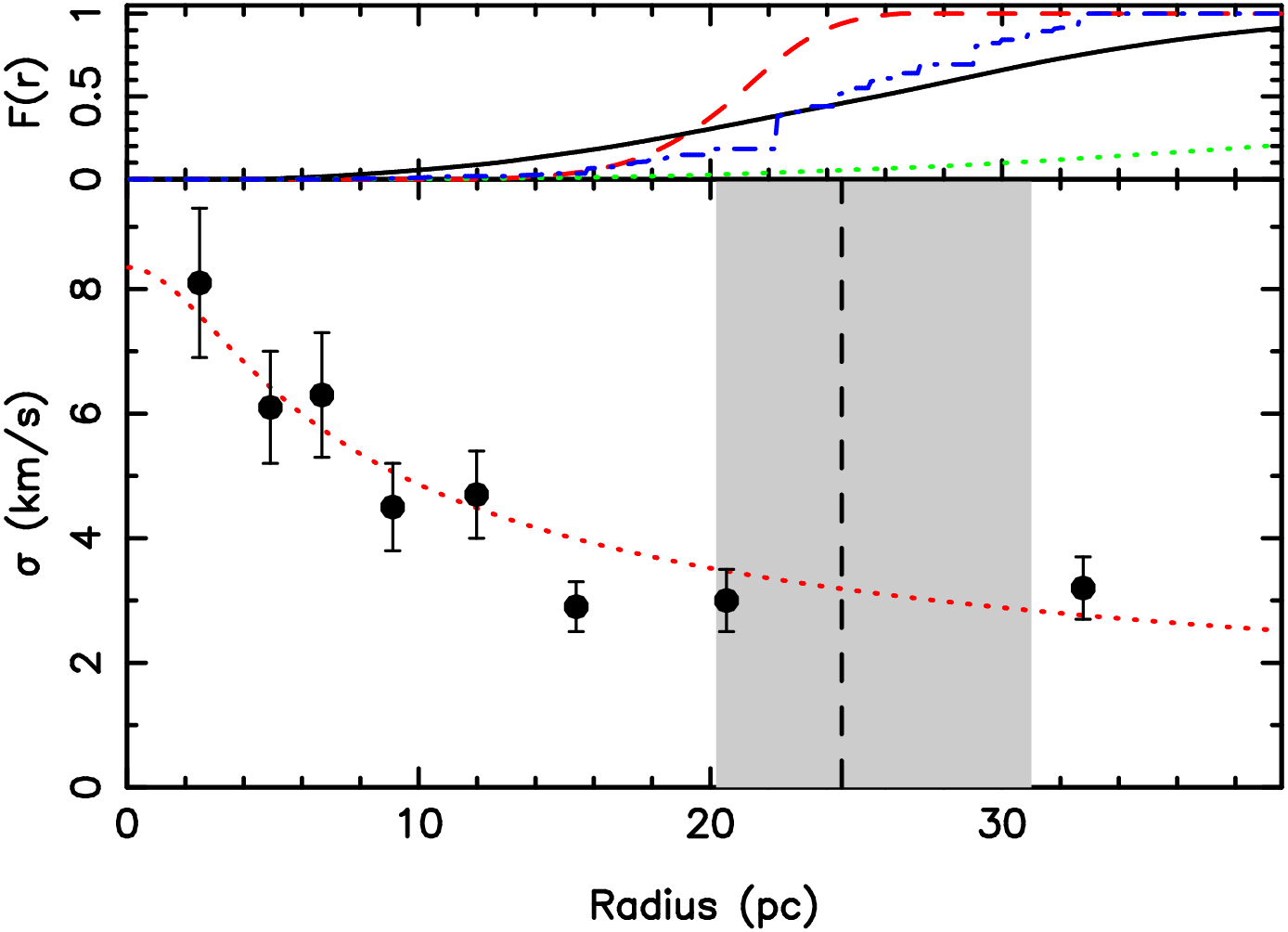} \\
\multicolumn{1}{c}{\mbox{}} \\
\multicolumn{1}{l}{\mbox{(e) {\bf NGC 7099}; $M_C$ = $1.12 \times 10^{5} \msun$}} \\[-0.1cm]
\includegraphics[width=\figsizeFour,angle=270]{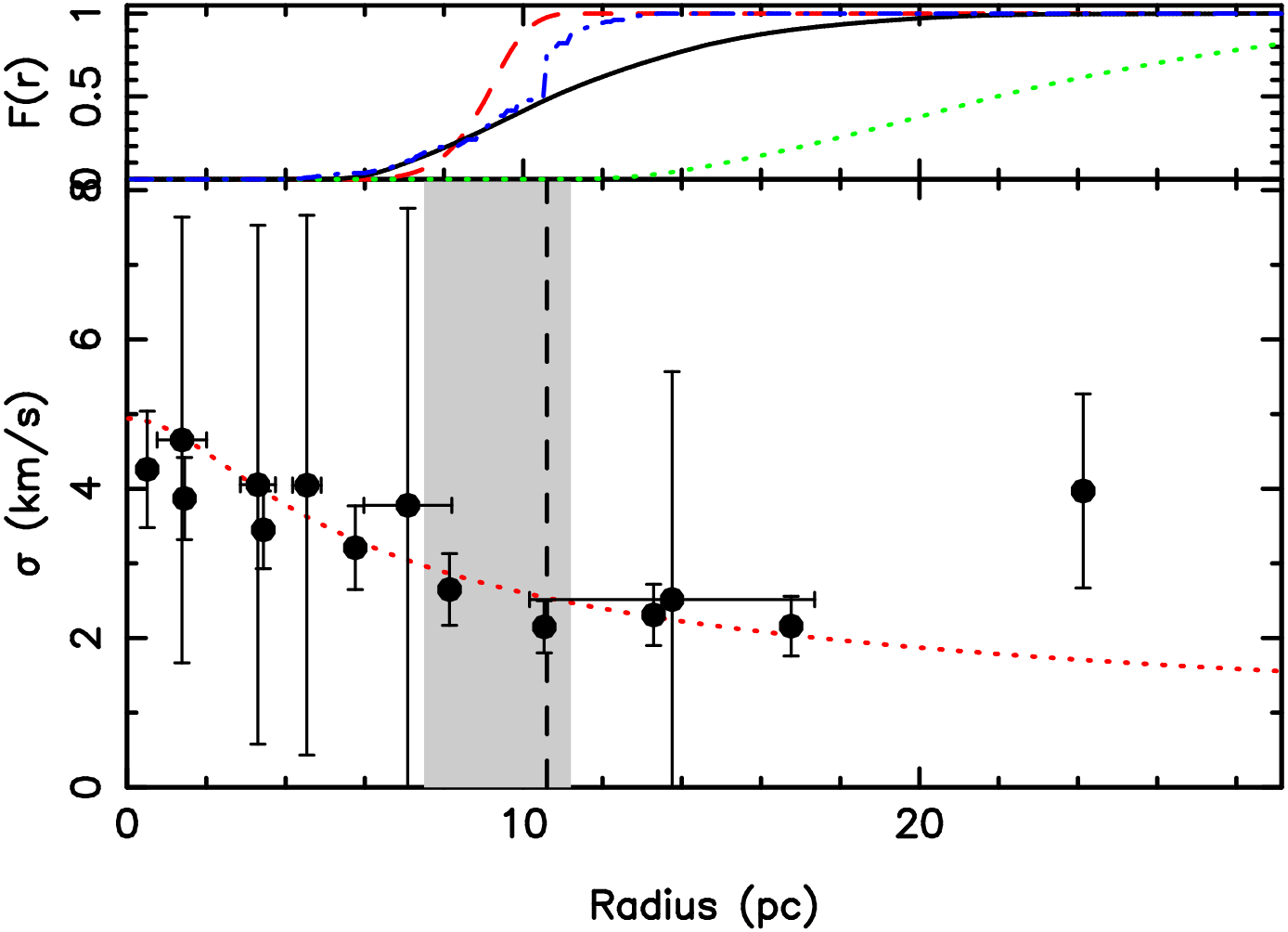} 
\end{array}$ \par\end{centering}
\caption{Continuation of Figure~\ref{VD_All_DIST1} and~\ref{VD_All_DIST2}.}
\label{VD_All_DIST3} 
\end{figure}

From Figures~\ref{VD_All_DIST1}-\ref{VD_All_DIST3} it is clear that there is a large range of values for $r_{c}^*$ and for $r_t$, but not quite as large a spread for $r_{m}$. This is due to the MOND model only depending on the cluster mass, while the chaos radius and tidal radius heavily depend on the GC-galaxy orbital eccentricity ($e$) and perigalacticon ($R_p$) which have large uncertainties. Changing either of these quantities has a dramatic effect on the estimate for $r_{c}$, for example an increase in eccentricity of $\Delta e \sim 0.1$ will produce approximately a 10\% decrease in $r_{c}$, for a fixed value of $R_p$. The observational uncertainties and their effect on the work presented here is discussed in the next section.

\subsection{Observational uncertainties and complications}

\label{obsuncert}

As previously stated the key requirements for predicting the flattening of GCs is the cluster mass and the GC-galaxy orbital parameters. Both of which have large observational uncertainties. 

Observationally determined GC masses are sensitive to assumptions made about stellar distribution and mass to light ratio. In Table~\ref{TblObs} the mean and standard deviations were shown from the different cluster masses found in the literature. Often each individual study quotes a cluster mass and observational errors which are not in agreement with other studies. For consistency between clusters we have used the velocity dispersion values close to the core of the cluster to approximate the central velocity dispersion and the half mass radius to determine the cluster mass from \reqOne{MCSigma}.

By using the central velocity dispersion to determine the mass the uncertainty in the distances to each cluster from the solar system is relevant. Any error in the distance will directly effect the half-mass radius during the conversion from arc-minutes to pc, and through \reqOne{MCSigma} the cluster mass. The relative errors in the distance are of the order 10\% which will have the same magnitude uncertainty on both $r_{h}$ and $M_C$. As seen in  Table~\ref{TblObs}, and particularly in Figure~\ref{fig:GCMassProb}, the uncertainties in the mass are typically larger than this and so the effect of the distance has been neglected.

Errors in the GC orbital parameters ($R_p$ and $e$) are primarily due the observed velocity components tangential to our line of sight having large uncertainties. Such measurements are difficult to obtain and rely on comparing the current positions of globular clusters on the sky with older position data, for example those found on photographic plates. The motion of each cluster is then determined from the change in positions (in milli-arcseconds) over time, with the positions calibrated using ``fixed'' distant objects, such as quasars \citep[e.g.][]{DinescuI}. The error is also amplified by uncertainties in the distance to the cluster ($R_G$). For the orbital parameters in Table~\ref{TblObs} these uncertainties are most likely to translate into large errors in eccentricity if $R_p/R_G \sim 1$ and in eccentricity and perigalacticon if $R_p << R_G$. 

The clusters NGC 1851 and NGC 7078, and to a lesser extent NGC 1904, all have high radial velocity components making their galactic orbits extremely radial and the determination of the apocentre particularly prone to uncertainties. In addition, NGC 7078 has a poorly measured absolute proper motion (Dinescu-Casetti, private communication) for which four discrepant determinations exist. For this cluster the proper motions of \cite{Odenkirchen1997} are used, but extra caution is advised regarding the orbital determination.

Another consequence of the poor determination of the proper motions is the effect on the orbital phase. The orbital phase for each cluster is given in Column 7 of Table~\ref{TblObs}, from which most clusters appear to be close to apocentre. Statistically all clusters are not likely to be so close to apocentre at the same time and the orbital phase values are most likely due to the bias in the orbital parameters towards radial orbits. This makes the orbital phases particularly unreliable, fortunately the predicted radii given in Table~\ref{TblRadii} are not sensitive to the phase.

The effect of the uncertainty in the orbital elements ($R_p$ and $e$) is included in these figures by distributing the velocities by their error bars and calculating 1000 different orbits for each cluster.  As seen in Table~\ref{TblObs} the relative error in both of these quantities is $\gapprox$ 10\%  and is by far the largest source of error comparing the different models to the flattening radii. The error in the orbital elements also exceeds any error due to the assumption of a point mass galactic potential, although this will be further investigated in an upcoming publication.

\section{Model comparison}

\label{Comparison}

The aim of this section is to compare the three models for the flattening of the velocity dispersion profile in GCs. The three models to the flattening radius are; (1) the corrected chaos radius $r_c^*$, (2) the MOND acceleration radius $r_{m}$ and (3) no flattening, i.e. using the original equilibrium velocity dispersion. The approach taken here is to use a variety of statistical analyses of the $10^6$ realisations for the GC-galaxy orbit ($R_p$ and $e$) and cluster mass ($M_C$) to answer some key questions. These key questions are: is flattening in the velocity dispersion profile required? which is the better flattening model, chaos or MOND? can these two models be easily distinguished? how good is the present observational data? can the models be used to constrain the cluster masses and orbital parameters?

To answer the first two questions a likelihood based approach is adopted and the likelihood ratios between models are sought. This requires the probability of the data given the model which we equate to the likelihood, i.e. $P(\sigma|M_i) = \mathcal{L}(M_i|\sigma)$. Assuming a normal distribution of the error $e_k$ associated with each velocity dispersion observation $\sigma_k$ at radius $r_k$ then the likelihood is given by
\begin{equation}
\mathcal{L}_j(M_i|\sigma) = 
\left( \frac{1}{2\pi \bar{e}^2} \right)^{n_{obs}/2} 
\exp \left( - X_i \right)
\label{EqnLikelihood}
\end{equation}
where $j$ refers to a single set of ($M_C$, $R_p$ and $e$) taken from the $10^6$ realisations per GC and 
\begin{equation}
X_i = \sum_{k=1}^{n_{obs}} \left( \frac{\sigma_k(r_k) - f(r_k|M_i)}{e_k \sqrt{2}} \right)^2
\label{eqnleastsquares}
\end{equation}
where $\bar{e}$ is the average velocity dispersion uncertainty and $n_{obs}$ is the total number of data points. The model $M_i$ refers to either no flattening (denoted by $N$) with $\sigma(r)$ given by \reqOne{SigmaR}, or flattening at the chaos radius ($C$), or at the MOND radius ($M$) or at the best fit flattening radius ($F$). 

Broadly speaking there are two ways to calculate the likelihood for each model; the first is to determine the likelihood $\mathcal{L}_j(M_i|\sigma)$ for each of the $j=1 \dots 10^6$ realisations and multiply them to get the final likelihood for each GC, i.e. 
\begin{equation}
\mathcal{L}(M_i|\sigma) = \prod_{j=1}^{10^6} \mathcal{L}_j(M_i|\sigma)
\end{equation}
the second approach is to use the maximum likelihood estimation across all $10^6$ realisations, i.e. 
\begin{equation}
\mathcal{L}(M_i|\sigma) = \sup\limits_{j} \left( \mathcal{L}_j(M_i|\sigma) \right) . 
\end{equation}
The second approach is used here since the first introduces a large bias for outlier orbits; e.g. if an individual likelihood $\mathcal{L}_j(M_i|\sigma) \sim 0$ then the overall likelihood also goes to zero, even if all other $10^6$ realisations give good fits to the data. This is particularly a problem for poorly constrained orbits and thus affects the chaos model more than the MOND model.

To directly compare the chaos and MOND models the maximum likelihood estimates are used to construct a likelihood ratio given by
\begin{equation}
K = \frac{P(\sigma|M_C)}{P(\sigma|M_M)} = \frac{\mathcal{L}(M_C|\sigma)}{\mathcal{L}(M_M|\sigma)} = \exp(X_M-X_C)
\end{equation}
where $X_i$ is defined by \reqOne{eqnleastsquares} and the average uncertainty $\bar{e}$ does not need to be defined as it falls out of the ratio. By construction the likelihood ratio $K$ is larger than unity when the chaos model is favoured by the data compared to the MOND model. To determine if flattening is required to model the data at all a similar likelihood ratio is constructed by
\begin{equation}
K_F = \frac{P(\sigma|M_F)}{P(\sigma|M_N)} = \frac{\mathcal{L}(M_F|\sigma)}{\mathcal{L}(M_N|\sigma)} = \exp(X_N-X_F)
\end{equation}
where model $N$ refers to the equilibrium model without any flattening and model $F$ to the best fit model with flattening. If $K \leq 1$ then no flattening is required to explain the observed velocity dispersion.

The likelihood ratio comparing the comparing the chaos and MOND models ($K$) and the best fit flattening to no flattening models ($K_F$) are shown in columns 10 and 11 of Table~\ref{GCSummary}. This table is ordered by decreasing $K_F$ so the GCs at the top have greater observational evidence for flattening. Adopting Jeffreys' interpretation for the strength of the comparison \citep{Jeffreys1998}, those with $K_F \lapprox 3$ can be fit without requiring any flattening, those with $K_F > 3$ show substantial evidence for requiring flattening and those with $K_F > 10$ have strong evidence. For the GCs NGC 104, NGC 5139 and NGC 6171 the likelihood ratio decisively supports flattening, which makes these clusters (especially NGC 6171) good candidates for distinguishing between models (see Section~\ref{disc:good}). However, as discussed in Section~\ref{disc:weak}, NGC 104 and NGC 5139 have issues which make the interpretation of the likelihood ratios more complicated.

\begin{table*}
\caption{Comparison between models for all 15 globular clusters examined in this work.}
\label{GCSummary}
\begin{centering}
\begin{tabular}{lcccccccccc}
\hline 
Cluster & $r_f/r_t$ & $\eta_{mass}$ & $\eta_{orb}$ & $\eta_{rot}$ & $\epsilon$ &  $D_C$ & $D_M$ & $D_{CM}$ & $K$ & $K_F$ \\
\hline
\vspace{\tablegap}
NGC  6171 &     0.57$^{+    0.31}_{-    0.18}$ &     0.16 &     0.09 & 0.71 &     0.02 &     0.33 &     0.55 &     0.60 &     5.38 &  4226.52 \\ \vspace{\tablegap}
NGC  5139 &     1.55$^{+    0.45}_{-    0.31}$ &     0.19 &     0.05 & 0.32 &     0.17 &     1.00 &     1.00 &     1.00 &     $2 \times 10^{-18}$ &   673.13 \\ \vspace{\tablegap}
NGC   104 &     0.24$^{+    0.12}_{-    0.08}$ &     0.25 &     0.01 & 0.46 &     0.09 &     0.83 &     0.83 &     0.38 &     3.35 &   116.39 \\ \vspace{\tablegap}
NGC  7099 &     0.43$^{+    0.16}_{-    0.13}$ &     0.18 &     0.09 & 0.00 &     0.01 &     0.34 &     0.49 &     0.50 &     1.30 &    17.41 \\ \vspace{\tablegap}
NGC  6341 &     0.90$^{+    0.17}_{-    0.23}$ &     0.17 &     0.07 &   0.36$^*$ &     0.10 &     0.86 &     0.50 &     0.97 &     4.05 &     6.28 \\ \vspace{\tablegap}
NGC  7078 &     0.42$^{+    0.26}_{-    0.12}$ &     0.25 &     0.11 & 0.28 &     0.05 &     0.25 &     0.52 &     0.49 &     1.25 &     5.37 \\ \vspace{\tablegap}
NGC  1851 &     1.45$^{+    0.54}_{-    0.59}$ &     0.02 &     0.17 & 0.15 &     0.05 &     1.00 &     1.00 &     1.00 &     2.09 &     4.65 \\ \vspace{\tablegap}
NGC  1904 &     1.13$^{+    0.87}_{-    0.50}$ &     0.25 &     0.17 & 0.11 &     0.01 &     0.82 &     0.88 &     0.66 &    81.31 &     2.96 \\ \vspace{\tablegap}
NGC   288 &     0.54$^{+    0.43}_{-    0.32}$ &     0.24 &     0.12 & 0.19 &    - &     0.35 &     0.77 &     0.92 &     2.99 &     1.73 \\ \vspace{\tablegap}
NGC  6121 &     1.90$^{+    0.10}_{-    0.71}$ &     0.29 &     0.09 & 0.46 &     0.00 &     0.94 &     0.58 &     1.00 &     1.27 &     1.68 \\ \vspace{\tablegap}
NGC  5024 &     0.12$^{+    0.08}_{-    0.03}$ &     0.24 &     0.14 & 0.00 &     0.01 &     0.93 &     0.27 &     0.97 &     1.03 &     1.19 \\ \vspace{\tablegap}
NGC  6656 &     0.21$^{+    0.16}_{-    0.17}$ &     0.34 &     0.03 & 0.22 &     0.14 &     0.76 &     0.82 &     0.60 &     1.08 &     1.10 \\ \vspace{\tablegap}
NGC  6218 &     1.10$^{+    0.51}_{-    0.34}$ &     0.12 &     0.10 & 0.06 &     0.04 &     0.95 &     0.92 &     1.00 &     1.03 &     1.03 \\ \vspace{\tablegap}
NGC  6809 &     0.23$^{+    0.76}_{-    0.23}$ &     0.36 &     0.05 & 0.19 &     0.02 &     0.69 &     0.46 &     0.49 &     1.01 &     1.00 \\ \vspace{\tablegap}
NGC  6752 &     0.21$^{+    0.36}_{-    0.10}$ &     0.31 &     0.02 & 0.00 &     0.04 &     0.80 &     0.64 &     0.60 &     1.08 &     0.99 \\
\hline
 \end{tabular}
\end{centering}

\medskip
Comparison between models for all GCs ordered by the strength of observational evidence favouring flattening ($K_F$) in their velocity dispersion profiles. Observationally derived measurements of the cluster mass error ($\eta_{mass}$), orbital uncertainties ($\eta_{orb}$), cluster rotation ($\eta_{rot}$) and cluster ellipticities ($\epsilon$) are shown in columns 3-6. Columns 7 and 8 give the K-S statistics for comparing the chaos and MOND distributions to the best fit flattening radius distribution. A measure of how easy the chaos and MOND distributions are to distinguish is given in column 9. The likelihood ratios in favour of the chaos model over the MOND model and for comparing the best fit flattening model to the no flattening model are given in columns 10 and 11 respectively. $^*$ The $\eta_{rot}$ value for NGC 6341 is unreliable, see in-text for details.
\end{table*}
%

Regarding the comparison between flattening at the chaos radius and flattening at the MOND radius most GCs in Table~\ref{GCSummary} have a likelihood ratio of $K \sim 1$. Four clusters favour the chaos radius (NGC 104, 6341 and 6171) one very strongly (NGC 1904) and one cluster (NGC 5139) strongly favours MOND. However for NGC 5139 neither the chaos or MOND models are consistent with the best fit flattening radius (see Figure~\ref{VD_All_DIST1} f). This cluster is discussed in more detail in Section~\ref{disc:weak}, for now we seek a method to compare the distribution functions for the flattening radius between models.

The Kolmogorov-Smirnov (K-S) statistic is used to test the similarities between two cumulative distribution functions. Here it is used firstly to compare the distribution for the chaos or MOND radius to the distribution function for the best fit flattening radius and secondly to compare the chaos and MOND distributions directly. In the first case it is defined as
\begin{equation}
D_X = \rm{sup} \left| F_x(r) - F_f(r) \right|
\end{equation}
where $X = C$ or $M$ and denotes either the cumulative distribution for $r_c^*$ (black curves in Figures~\ref{VD_All_DIST1}-\ref{VD_All_DIST3}) or $r_m$ (red dashed curves). Both of which are compared to the empirical cumulative distribution function for the best flattening fit $r_f$, shown as blue dot-dashed curves in the figures. A value of $D_X = 0$ indicates that model $X$ is drawn from the same distribution as the best flattening fit distribution, conversely $D_X = 1$ indicates they are drawn from distinct distributions. The resulting K-S statistics for each cluster are given in Table~\ref{GCSummary} for the chaos model ($D_C$ in column 7) and the MOND model ($D_M$ in column 8).

The second K-S statistic comparing the cumulative distribution for $r_c^*$ directly to the distribution for $r_m$ is shown in column 9 of Table~\ref{GCSummary}. This statistic typically has $D_{CM} \sim 1$ which means that the distributions are easy to distinguish for all clusters except NGC 104 and NGC 7078 (depending on the orbit). This answers the third of the key questions in that generally the chaos and MOND models can easily be distinguished. 
Note that all $D_X$ values in Table~\ref{GCSummary} are greater than the critical value of $D_{\alpha=0.05} = 0.04$; determined using $N=10^3$ independent sets of orbital parameters. The $D$ values are quoted here since the associated K-S probabilities \citep[calculated using the method given in ][]{AstroStats2012} are all $\lapprox 10^{-20}$  which indicates that all radii distributions are drawn from different underlying distribution functions. 

In addition to model comparison statistics, Table~\ref{GCSummary} lists observationally derived measurements of the distribution of ratios between the best fit flattening radius and the tidal radius (column 2), cluster mass error (column 3), orbital uncertainties (column 4), cluster rotation (column 5) and cluster ellipticities (column 6). The ratio between the flattening and tidal radius ($r_f/r_t$) is a simple measure of the ease of observing any deviation from an equilibrium velocity dispersion model, where high values imply that it is very difficult to disentangle any flattening from classical tidal truncation.

To gauge the cluster mass error, $\eta_{mass}$ is defined as the maximum error in the fitted masses from column 3 of Table~\ref{TblObs} over the median cluster mass. The uncertainty for the galactic orbit of each cluster ($\eta_{orb}$) is defined as an average of the positive and negative relative errors for $R_P$ and $R_A$. High values of $\eta_{orb}$ indicate large uncertainties in the orbital parameters of the cluster. As seen from Table~\ref{GCSummary} most clusters have relative errors of $\eta_{mass}$ $\sim$ 20\% and $\eta_{orb}$ $\sim$ 10\%. Both of these affect the results, with errors in the orbit affecting the tidal and chaos radii more, while the MOND radius is only affected by the cluster mass. Overall the orbital uncertainty is the dominant source of error affecting this study.

The relative importance of rotation for each cluster is defined here as the ratio between the magnitude of the rotational velocity and the central velocity dispersion value ($\sigma_0$). The rotational velocity is assumed to be of the form $A_{rot} \sin (x)$ where $x$ depends on the position angle inside the cluster; here only the magnitude $A_{rot}$ is of interest. The ratio $\eta_{rot} = A_{rot}/\sigma_0$ is given in column 5 of Table~\ref{GCSummary} for all GCs with values for $A_{rot}$ taken from \cite{BellazziniEtAl2012} except NGC 6341 which is approximated from figure 12 of \cite{DCLSBM2007} ($A_{rot} \sim 2.5$ km/s and $\sigma_0 \sim 7$ km/s). Rotation in GCs will cause the cluster to become oblate, which means that the equilibrium velocity dispersion profile given by \reqOne{SigmaR} is invalid as it assumes spherical symmetry. The observed ellipticities from  \cite{Harris1996} (2010 edition) are given in column 6 of Table~\ref{GCSummary} and generally follow the same trend as $\eta_{rot}$.

Of the key questions posed previously the likelihood ratios can answer if flattening is required ($K_F$) and if the chaos or MOND model is prefered ($K$) while the K-S statistic shows when these models can be distinguished ($D_{CM}$). For the remainder of this section we focus on the quality of the velocity dispersion observational data and if the models examined here can be used to constrain observational parameters.

To examine how well the data fits each model we test to see if the measurement error alone can account for the scatter seen in the velocity dispersion. This is achieved by using the coefficient of determination defined as
\begin{equation}
R^2_{i,j} = 1 - \frac{ \sum_{k=1}^{n_{obs}} \left( \sigma_k(r_k) - f(r_k|M_i) \right)^2 }
{ \sum_{k=1}^{n_{obs}} \left( \sigma_k(r_k) - \mu \right)^2 }
\end{equation}
where $i$ denotes the type of model ($C$, $M$, $\dots$) and $j$ is a single set of cluster mass and orbital parameters of the $10^6$ total realisations. The mean velocity dispersion value is
\begin{equation}
\mu = \frac{\sum_{k=1}^{n_{obs}} \sigma_k(r_k) }{n_{obs}}.
\end{equation}
A successful model has $R^2_{i,j} \sim 1$, but can be negative if the model fit is worse than a horizontal line at the mean velocity dispersion value ($\mu$). The maximum $R^2_{i,j}$ values for each cluster are shown in Table~\ref{GCRSquared} for the best fit ($R^{2*}_F$), no flattening ($R^{2*}_N$), MOND ($R^{2*}_M$) and chaos ($R^{2*}_C$) models. Since each cluster has $10^6$ realisations then there is a range of $R^2_{i,j} \leq R^{2*}_{i}$ for each model $i$. The maximum coefficient of determinations for the best fit model ($R^{2*}_{F}$) are used to categorise each cluster (see below).

\begin{table*}
\caption{Maximum R-squared values and best orbital parameters for all 15 globular clusters.}
\label{GCRSquared}
\begin{centering}
\begin{tabular}{lcccccccccc}
\hline 
Cluster & $R^{2*}_{F}$ & $R^{2*}_{N}$ & $R^{2*}_{M}$ & $R^{2*}_{C}$ & $\widehat{M}_{C,F}$ & $\widehat{M}_{C,N}$ & $\widehat{M}_{C,M}$ & $\widehat{M}_{C,C}$ & $\widehat{R}_{p,C}$ & $\widehat{e}_C$  \\
        &           &  &  &  & ($\times 10^5 \msun$) & ($\times 10^5 \msun$) & ($\times 10^5 \msun$) & ($\times 10^5 \msun$) & (kpc) &   \\
\hline
\vspace{\tablegap}
NGC  6171 &     0.97 &     0.46 &     0.88 &     0.97 & $    0.93\pm    0.05$ & $    1.00\pm    0.05$ & $    0.95\pm    0.05$ & $    0.93\pm    0.05$ & $    2.15\pm    0.26$ & $    0.24\pm    0.06$ \\ \vspace{\tablegap}
NGC  1851 &     0.90 &     0.83 &     0.86 &     0.90 & $    3.73\pm    0.07$ & $    3.75\pm    0.06$ & $    3.74\pm    0.07$ & $    3.73\pm    0.07$ & $    3.53\pm    0.36$ & $    0.78\pm    0.04$ \\ \vspace{\tablegap}
NGC  6218 &     0.90 &     0.82 &     0.82 &     0.90 & $    1.04\pm    0.11$ & $    1.04\pm    0.11$ & $    1.04\pm    0.11$ & $    1.03\pm    0.11$ & $    1.12\pm    0.40$ & $    0.66\pm    0.10$ \\ \vspace{\tablegap}
NGC  7078 &     0.85 &     0.85 &     0.84 &     0.85 & $    3.34\pm    0.27$ & $    3.46\pm    0.26$ & $    3.32\pm    0.27$ & $    3.36\pm    0.26$ & $    5.63\pm    1.19$ & $    0.59\pm    0.09$ \\ \vspace{\tablegap}
NGC  1904 &     0.85 &     0.79 &     0.73 &     0.86 & $    1.33\pm    0.11$ & $    1.36\pm    0.11$ & $    1.22\pm    0.13$ & $    1.32\pm    0.12$ & $    5.86\pm    1.08$ & $    0.59\pm    0.06$ \\ \vspace{\tablegap}
NGC  6341 &     0.84 &     0.78 &     0.80 &     0.84 & $    1.36\pm    0.05$ & $    1.38\pm    0.05$ & $    1.37\pm    0.05$ & $    1.36\pm    0.05$ & $    2.62\pm    0.34$ & $    0.60\pm    0.04$ \\ \vspace{\tablegap}
NGC  5139 &     0.80 &     0.76 &     0.76 &     0.55 & $   24.58\pm    0.77$ & $   25.12\pm    0.76$ & $   25.12\pm    0.76$ & $   20.88\pm    0.78$ & $    1.60\pm    0.01$ & $    0.60\pm    0.00$ \\ \vspace{\tablegap}
NGC  6752 &     0.50 &     0.52 &     0.49 &     0.52 & $    1.86\pm    0.41$ & $    2.01\pm    0.33$ & $    1.99\pm    0.34$ & $    2.00\pm    0.33$ & $    4.15\pm    0.36$ & $    0.13\pm    0.03$ \\ \vspace{\tablegap}
NGC  7099 &     0.41 &     0.11 &     0.41 &     0.41 & $    0.86\pm    0.07$ & $    0.91\pm    0.06$ & $    0.85\pm    0.07$ & $    0.86\pm    0.07$ & $    3.47\pm    0.57$ & $    0.35\pm    0.06$ \\ \vspace{\tablegap}
NGC   104 &     0.40 &     0.07 &     0.29 &     0.37 & $    9.52\pm    1.06$ & $   10.68\pm    0.72$ & $   10.49\pm    0.73$ & $   10.45\pm    0.73$ & $    4.09\pm    0.17$ & $    0.32\pm    0.03$ \\ \vspace{\tablegap}
NGC  6656 &     0.17 &     0.02 &     0.04 &     0.17 & $    3.24\pm    0.90$ & $    3.60\pm    0.76$ & $    3.57\pm    0.77$ & $    3.56\pm    0.77$ & $    3.07\pm    0.32$ & $    0.46\pm    0.03$ \\ \vspace{\tablegap}
NGC  5024 &     0.16 &    -1.16 &    -0.02 &     0.17 & $    4.97\pm    1.14$ & $    5.18\pm    1.07$ & $    4.99\pm    1.12$ & $    5.17\pm    1.08$ & $   14.26\pm    4.08$ & $    0.34\pm    0.13$ \\ \vspace{\tablegap}
NGC  6121 &     0.00 &    -2.32 &    -1.06 &     0.02 & $    1.43\pm    0.40$ & $    1.61\pm    0.34$ & $    1.54\pm    0.35$ & $    1.17\pm    0.29$ & $    0.63\pm    0.18$ & $    0.80\pm    0.06$ \\ \vspace{\tablegap}
NGC  6809 &    -0.11 &    -0.10 &    -0.15 &    -0.12 & $    0.93\pm    0.22$ & $    0.97\pm    0.21$ & $    0.94\pm    0.23$ & $    0.95\pm    0.21$ & $    1.77\pm    0.31$ & $    0.52\pm    0.08$ \\ \vspace{\tablegap}
NGC   288 &    -0.16 &    -0.37 &    -0.04 &    -0.15 & $    0.85\pm    0.12$ & $    0.91\pm    0.08$ & $    0.60\pm    0.05$ & $    0.85\pm    0.08$ & $    3.38\pm    1.22$ & $    0.57\pm    0.11$ \\
\end{tabular}
\end{centering}

\medskip
Values for the coefficient of determination for best fit flattening ($R^{2*}_{F}$; used for sorting), no flattening ($R^{2*}_{N}$), MOND ($R^{2*}_{M}$) and chaos ($R^{2*}_{C}$) models. The expectation values for the cluster masses and associated errors using these models are given in columns 6-9. For the chaos model constraints can be put on the orbital parameters $R_p$ and $e$ which are shown in columns 10 and 11 and can be compared to future orbital determinations when better proper motion observations become available.
\end{table*}

By considering all $10^6$ realisations, and appropriately weighting them, then expectation values can be made for the cluster mass $M_C$ for each model (and $R_p$, $e$ for the chaos model). The weight given to a particular set of parameters $j$ for a given model $i$ is
\begin{equation}
\omega_{i,j} = \frac{P(\sigma|M_{i,j})}{ \sum_{k=1}^{10^6} P(\sigma|M_{i,k})}
\label{weights}
\end{equation}
where the probabilities of each model, including the set of orbital parameters, $P(\sigma|M_{i,j})$ has previously been given by \reqOne{EqnLikelihood}. The weighted mean for the cluster mass is then found for each model $i$ by summing over all realisations $j$
\begin{equation}
\widehat{M}_{C,i} = \sum_{j=1} \omega_{i,j} M_C(M_{i,j})
\label{weightsmean}
\end{equation}
where $M_C(M_{i,j})$ is the cluster mass used as an input parameter for $M_{i,j}$. A similar equation can be written for the standard deviation which is used here to estimate the error for the expected cluster mass. This process is applied to the cluster masses for each model (best fit, no flattening, flattening at chaos radius and flattening at the MOND radius) and the resulting estimated values and errors are given in Table~\ref{GCRSquared}. For the chaos model the observational parameters ($R_p$ and $e$) can also be estimated by modifying \reqOne{weightsmean}, the expected orbits are shown in columns 10 and 11 of Table~\ref{GCRSquared}.

The estimates for the cluster masses given in Table~\ref{GCRSquared} generally agree between models and with the literature values given in Table~\ref{TblObs}. The exception to this is NGC 5139 where all estimates are lower than both the literature values and the best fit to the central velocity dispersion. The NGC 5139 estimation based on the chaos model is lower again, however this is a weak fit since the fit to the data for this model is poor (e.g. $R^{2*}_{C} = 0.55$).

The orbital parameters estimated by assuming the chaos model is correct are given in columns 10 and 11 of Table~\ref{GCRSquared}, these are also consistent with Table~\ref{TblObs} for most clusters. The estimated orbits give more realistic (i.e. less radial orbits) $R_p$ and $e$ values for NGC 1851, NGC 1904 and marginally better for NGC 5139 and NGC 6341. The chaos model predictions are of great interest since these can be compared to future GC orbital determinations when better proper motion data is available, e.g. after GAIA or using tidal tail observations.

Results from the comparison between models shown in Tables~\ref{GCSummary} and~\ref{GCRSquared} are divided into three broad categories; (1) clusters undergoing disruption, (2) clusters with too weak evidence or unknown origin such as NGC 5139, and (3) candidate systems to distinguish between the chaos and MOND models. Since these tables are effectively a summary of this paper the results are discussed in detail along with a more general discussion in the next section.

\section{Discussion}

\label{IndivGCs}

The aim of this section is two-fold; firstly to discuss the results from Tables~\ref{GCSummary} and~\ref{GCRSquared} by dividing the GCs into categories, and secondly to narrow down the list of GCs to provide the most promising candidates for further observations.

Generally all GCs have a likelihood ratio $K>1$ and so there is no probabilistic motivation to throw out Newton (chaos model) in favour of Milgrom (MOND model). To divide the GCs into meaningful groups the $R_F^2$ values from Table~\ref{GCRSquared} are used\footnote{For convenience the * is dropped from best coefficient of determination notation, i.e. $R^2_F = R_F^{2*}$.} and it is seen that there are no clusters with $0.5 < R_F^2 < 0.8$, so the first division is taken as $R_F^2 = 0.6$. For the GCs with $R_F^2 < 0.6$ a further division can be made using $K_F \sim 3$, with clusters above this divide favouring flattening. These divisions result in the same categories mentioned in the last section, namely (1) clusters undergoing disruption, $R_F^2 < 0.6$ and $K_F \lapprox 3$, (2) clusters with weak evidence, $R_F^2 < 0.6$ and $K_F \gapprox 3$ or GCs with unknown origin, i.e. NGC 5139, and (3) candidate systems to distinguish between Newtonian and MOND models, $R_F^2 > 0.6$. These categories are discussed in turn and finally complications to the model and the approach taken here are also discussed.

\subsection{Clusters undergoing disruption}

\label{disc:disrupt}

This category is defined as those GCs whose observed velocity dispersion profile is consistent with ongoing tidal disruption. Such clusters are characterised by low, and sometimes negative, $R^2_F$ values, i.e. their velocity dispersion profiles are better fit by a flat profile at the mean velocity dispersion. Specifically GCs with $R^2_F < 0.6$ and $K_F \lapprox 3$, these conditions are satisfied by NGC 288, NGC 5024, NGC 6121, NGC 6656 and NGC 6809. 

Three of these clusters, NGC 288, NGC 6121 and NGC 6809 have their predicted chaos radii very close to their observed half-mass radii. At present the velocity dispersion profile around the half-mass radius of NGC 6121 is unresolved ($r_h = 3.3\pc$) however the profiles of NGC 288 and NGC 6809 are quite flat even within their half-mass radii. Both of these clusters have low masses ($M_C \sim 9 \times 10^4 \msun$) with half-mass radii of 6.9 pc and 5.3 pc respectively, thus they have very low concentrations. For this reason these three clusters are thought to be undergoing tidal disruption.

The MOND model generally predicts NGC 288 and NGC 6809 to be flat since the internal cluster accelerations are below the MOND acceleration for all radii. However both GCs spend most of their orbits too deep in the galactic potential for MOND to be applicable, respectively only 38\% and 28\% of their orbital time is spent with sufficiently low acceleration. The chaos and tidal radii are difficult to determine for these clusters due to observational uncertainty in the orbital parameters, but they are typically close to the half-mass radii, which implies that both clusters are being disrupted. Evidence of ongoing disruption of these clusters is seen in tidal tail observations for NGC 288 \citep{GrillmairEtAl1995,LeonEtAl2000} and NGC 6809 \citep{LeonEtAl2000}.

\citet{LaneEtAl2011} found that NGC 6121 had a mass of twice the literature values due to tidal heating increasing the velocity dispersion in the outer regions. The velocity distribution of this cluster was also complicated by the clear signature of cluster rotation found in the observations. As yet, no tidal tail studies were found in the literature for NGC 6121. More velocity dispersion observations are needed for this cluster to determine if it too is being disrupted as is predicted.

Three more clusters have low $R^2_F$ values, consistent with a flat velocity dispersion, these being NGC 5024, NGC 6656 and NGC 6752. These are more likely to be due to the very poor observational data currently available for the velocity dispersion.

\subsection{Weak evidence or unknown origin}

\label{disc:weak}

GCs with weak evidence are defined as where the case for flattening of the velocity dispersion profile depends on a single data point (typically with large error bars) at a distance that is likely to be beyond the tidal radius. Two GCs clearly in this category are NGC 104 and NGC 7099 which both satisfy $K_f > 10$ and $R^2_F < 0.6$. Also included in this category is NGC 5139 (despite having $R^2_F = 0.8$) since it does not appear to truly belong in the GC sample.

As is typical for this cluster, the velocity dispersion profile of NGC 5139 ($\omega$ Cen) presents difficulties for all models. The MOND model has particular difficulties with this cluster for two reasons; firstly the acceleration due to the galactic potential is never below the MOND limit, secondly the cluster mass is very large which leads to very large values of $r_{m}$. Explaining the possible flattening seen around 30 pc is also difficult for the chaotic orbits model as the stability boundary is found to be well inside the cluster for the orbital parameters given in Table~\ref{TblObs}. Two possible resolutions to this incompatibility might come from the fact that NGC 5139 is rapidly rotating \citep[e.g.][]{SF2010} and/or that it is suffering ongoing disruption. There is evidence of ongoing tidal disruption in that tidal tails have been observed for NGC 5139 \citep{LeonEtAl2000}.

Note that the likelihood ratio comparing the chaos to MOND models gives a decisive ruling in favour of MOND ($K = 2 \times 10^{-18}$). However this is not due to MOND giving a good fit to NGC 5139, but the chaos model giving a very poor fit. In fact MOND predicts $r_m = 60 \pc$ which is further from the cluster centre than all velocity dispersion data, i.e. MOND is equivalent to the equilibrium model for this cluster. This is seen in the identical $R^2_M$ and $R^2_N$ values in Table~\ref{GCRSquared}. Again the question of whether NGC 5139 is a true GC arises, answering this is beyond the scope of this paper.

For NGC 104 the final data point is at $r \sim 100\pc$, which is beyond the tidal radius, which has a median of $r_t = 57 \pc$ for all orbit realisations. For NGC 7099 the final data point is at $r \sim 24\pc$ and the median tidal radius is $r_t = 22 \pc$. For both of these clusters the main evidence for flattening comes from the single data point with $r \gapprox r_t$ where it is expected to contain significant contamination from the cloud of unbound stars remaining close to the GC, as discussed in Section~\ref{cloud}. 

A counter argument for NGC 104 comes from the observation that, at present, there are no statistically significant evidence of tidal tails \citep{LeonEtAl2000,LaneEtAl2010}. However these studies were examining stars that are clearly bound to the GC and so are unlikely to have seen the tidal tails. In the similar case of NGC 7099 tidal tails have been observed \citep{ChunEtAl2010}.

For NGC 7099 only one data point with large error bars deviates from the equilibrium values, furthermore this point only appears in the data from \citet{ScarpaEtAl2007} but not in \citet{LaneEtAl2011}. To resolve this discrepancy more accurate velocity dispersion observations in the outer region of the cluster are required. Given that NGC 7099 is non-rotating and has reasonably small uncertainties for the cluster mass and orbital parameters this cluster could be a potential candidate with more velocity dispersion observations at high radius. Clearer candidates for distinguishing between Newtonian and MOND model are discussed in the next section.

\subsection{Candidate systems}

\label{disc:good}

For the remaining 6 clusters, with $R_F^2 > 0.8$, the aim here is to provide the most promising candidates for further observations. A GC is promising if it satisfies the following criteria: (1) small $r_f/r_t$, as this is easier to observe, (2) large $D_{CM}$ so Newton and MOND can be distinguished (true for all 6 GCs), (3) strong evidence for flattening, i.e. $K_F > 3$, and (4) a minimal amount of additional observations needed to improve the orbit and velocity dispersion data. 

Using these criteria NGC 1904 and NGC 6218 can be eliminated as both may not need flattening at all ($K_F < 3$) and if they do it is consistent with the tidal radius (i.e. $r_f \sim r_t$). Also the present velocity dispersion errors are large for NGC 6218 and more resolution in radius is required for both clusters.

As discussed in Section~\ref{obsuncert}, both NGC 1851 ($e \sim 0.92$) and NGC 7078 ($e \sim 0.60$) are problematical clusters where large errors in proper motions are combined with high radial velocities, resulting in the integrated orbits being more radial than their true orbits. Large eccentricities have a strong effect on the chaos radius and on the tidal radius, so until the orbits are more accurately determined these predicted radii are very unreliable. 

In addition to poor orbital determination, the observational velocity dispersion data used for NGC 7078 is based on the no rotation model favoured by \citet{DSCLBMS1998}. They also put forward the explanation of the flattening in the velocity dispersion as being due to tidal heating, as expected in the models of \citet{AR1988}.  The difficulty in modelling NGC 7078 in the inner regions is also reflected in its high mass error, which is unsurprising given its more recent high rotation value given in Table~\ref{GCSummary}. The flattening in NGC 1851 is also likely to be due to tidal heating since the best fit flattening radius is consistent with the tidal radius. These factors make both clusters poor candidates at present until more orbital data becomes available.

The remaining clusters, NGC 6171 ($K_F \approx 4000$, $K \approx 5$, and $r_f/r_t = 0.6$) and NGC 6341 ($K_F \approx 6$, $K \approx 4$, and $r_f/r_t = 0.9$), are promising candidates for distinguishing Newtonian and MOND models since these clusters have small relative errors for both the cluster mass and orbital parameters. For NGC 6341 the existing data is quite good, although further observations of the velocity dispersion at large radii would better probe the region close to the tidal radius. In the case of NGC 6171 more resolution in radius is required (only 5 data points as seen in Figure~\ref{VD_All_DIST2} b), although the data coverage of the flattening region and the small error bars make this cluster a very good candidate.

Both clusters are rotating strongly enough that velocity dispersion inside the half-mass radius must be modelled in more detail. Interestingly increased rotation is predicted by the chaos diffusion model \citep{PaperI}. Specifically, the region between the chaos radius ($r_c$) and the maximum radius ($f_{max}$ in Figure~\ref{fig:rtidal}) will have a different rotational profile compared to the rest of the cluster. This is due to preferential removal of stars on prograde orbits relative to the GC-galaxy orbit compared to stars on retrograde orbits. Thus more stars on retrograde orbits will exist in the outer regions on the cluster, leading to a net rotation in this region. The long term effect of this rotation on the centre of the cluster, where the $\eta_{rot}$ values are measured, will depend on detailed cluster modelling. 

To answer questions regarding the emergence of rotation and the timescale for stars to escape via chaotic diffusion, an N-body simulation based on the mass and orbital parameters of NGC 6341 is currently in progress, the results from which will be presented in an upcoming publication.

NGC 6171 may already be showing evidence of chaotic diffusion of stars leading to flattening at the chaos radius. Clarification of this phenomenon will require more resolution in the velocity dispersion and better orbital determination. At present the data shows a clear need for a flattened model ($K_F = 4 \times 10^5$) but not so flattened as to be consistent with a completely flat profile ($R^2_F = 0.97$). In additional the flattening is observed well inside the tidal radius ($r_f/r_t = 0.6$) so it should be possible to see the flattening more clearly. Using the available data this cluster supports the Newton model over the MOND model with likelihood ratio of $K=5.4$. However it is the most rapidly rotating cluster in this sample with $\eta = 0.71$ so this should be taken into account more carefully before MOND can be ruled out entirely.

\subsection{Complications to models}

In the previous sections two additional complications to this study have emerged, namely rotation and tidal shocking. Previously known complications are the observational difficulties in determining the GC-galaxy orbit and the mass of the cluster.

Strong rotation seen in many clusters means that it is insufficient to use the velocity dispersion equilibrium model given by \reqOne{SigmaR}. As mentioned in Section~\ref{SimplePlummerCluster} the central velocity dispersion value, and therefore the cluster mass in our analysis, can change by up to 10\% for rotating clusters compared to non-rotating clusters \citep{EinselSpurzem1999,BaumgardtHutHeggie2002}. For rapidly rotating clusters the flattening models examined here can not be distinguished properly until rotation is included in the equilibrium model. However as mentioned in the context of NGC 6171 and NGC 6341 rotation may be enhanced by the preferential loss of prograde stars via chaos diffusion, more modelling is needed to test this.

Rotation complicates cluster mass determinations based on the central velocity dispersion, but as seen in Table~\ref{TblObs} there is already significant variation in the fitted cluster mass and values from the literature. The other major difficulty with this study is large uncertainties of the GC-galaxy orbits, described in detail in Section~\ref{obsuncert}. These two uncertainties have different effects on the flattening models as MOND is only sensitive to the cluster mass, while the chaos and tidal radii are more sensitive to the galactic orbit. The orbit can potentially be better determined using tidal tails.

The tidal field acting on a cluster close to the galactic disk and/or bulge will change over time, which causes the cluster to undergo tidal distortion \citep{OstrikerEtAl1972,OBS1989,KundicOstriker1995,GLO1999}. These two tidal processes are collectively referred to as tidal shocking and act to increase the mass loss of the cluster. \cite{WebbEtAl2013} found that tidal shocking can affect the energy of orbits to a depth of $\sim 0.14 r_t$ depending on the GC orbit. Distinguishing between tidal shocking and three-body instability, which both effect the orbital binding energy, is beyond the scope of the present paper. It will be examined in more detail in the third paper in this series.

Of the clusters listed in Table~\ref{TblObs} tidal tails have been observed in NGC 288, NGC 1851, NGC 1904, 5024, NGC 5139, NGC 6341, NGC 6809, NGC 7078, NGC 7099 \citep{GrillmairEtAl1995,LehmannScholz1997,LeonEtAl2000,Testa2000,ChunEtAl2010,JordiGrebel2010} possible indication of tidal tails for NGC 6218 \citep{LehmannScholz1997} no statistically significant evidence of tidal tails for NGC 104 \citep{LeonEtAl2000,LaneEtAl2010} and no tidal tail studies were found in the literature for NGC 6121, NGC 6171, NGC 6656 and NGC 6752. Evidence of tidal tails implies that these clusters are presently undergoing mass loss over the tidal radius, although determining the responsible process is observationally difficult. \cite{MontuoriEtAl2007} point out that cluster orbits from tidal tail observations are only accurate if stellar data are obtained from greater than 7-8 tidal radii from the GC centre. With this limitation in mind, further observations of tidal tails will substantially reduce the observational uncertainties of GC-galaxy orbits.

All clusters will lose mass over their lifetimes and the current population of GCs will be very different from the initial population \citep[e.g.][]{AHO1988,GO1997}. In other words the observed masses and orbital parameters are only their present values. Applying the formula by \cite{McLFall2008} to the half mass cluster densities for the 15 clusters here estimates that the initial masses were on average 3 times the present mass (up to 5 times for some clusters). An example of how this affects this analysis is via the relaxation times used in Table~\ref{TblRadii}. Originally the cluster would have more stars, so the relaxation timescale would be longer, and the escape timescale would be significantly shorter. Thus chaotic diffusion would have been even more efficient at removing stars from the outer regions of the cluster in the past. The orbit itself is not so sensitive to the mass of the cluster. Past orbital variations are accounted for using the method of back integrating the currently observed position and velocities for each cluster and then averaging over the previous 10 galactic orbits to get the parameters $R_p$ and $e$.

Recall from Section~\ref{theory:time} that both the tidal radius and the stability boundary have associated timescales for stars to escape or random walk from the inner to the outer parts of the cluster. Therefore there will be a finite time for the flattening of the radius to decrease to the predicted flattening radius. Since the cluster is continually losing mass, and its orbit around the galaxy is also changing, then the flattening radius is a moving target. So the observed flattening radius is always expected to be greater than the predicted flattening radius for the chaotic orbits and the tidal radius with the lag dependent on the response timescale, and therefore on the parameters of the GC-galaxy orbit. Note that this may also affect the predicted MOND acceleration radius, as it is unclear the timescale at which MOND would cause flattening of the velocity dispersion profile.

MOND also faces another problem when applied to the galactic GC system, namely the acceleration criterion. MOND is constructed to apply only when the acceleration on a body is less than $a_0 = 1.2 \times 10^{-10}$ m/s$^2$. Previous studies favouring MOND have ignored the acceleration acting on stars from the galactic potential, assuming that only the cluster potential contributes. This is not true and the fraction of time where each GC is in the MOND regime is given in Table~\ref{TblObs}. For many GCs this fraction is zero for all cluster masses and orbits, and so MOND will not apply. For more moderate fractions the validity of MOND for the cluster depends on the MOND timescale.

To summarise, more sophisticated modelling is needed which includes the precise orbit, inner cluster dynamics (especially rotation) and mass loss history of the cluster. There is no immediate requirement for such a model as the present observational uncertainties for the orbits dominate over errors introduced by an over-simplified model. Any improved model will require knowledge of the escape timescale for stars on unstable orbits and of the effect of non-Keplerian orbits on the stability boundary. Work in this direction using high resolution N-body simulations will be presented in the third paper in this series.

\section{Conclusions}
\label{Conc}

It was found that flattening of the velocity dispersion profiles for the 15 GCs investigated here can be explained using Newtonian dynamics models. To explain the flattening within a Newtonian framework the radius associated with the boundary between stars on stable and unstable orbits inside the GC was sought. Stars on unstable orbits will eventually escape the cluster potential through a process of chaotic diffusion.

The method used for determining the stability boundary is described in detail in the previous paper in this series \citep{PaperI} which was based on the stability of the general three-body problem as derived in \citet{RoIoA} and \citet{RoNew}. The predicted chaos radius was compared to the radius at which the acceleration drops below the critical MOND acceleration, which various authors have used to explain the observed flattening in the velocity dispersion profiles. A likelihood based analysis found that; firstly flattening is not required at all in some GCs and secondly for GCs with significant evidence of flattening there is no probabilistic motivation to throw out Newton in favour of Milgrom. 

As the stability boundary is a strong function of the parameters of the GC-galaxy orbit a detailed analysis of the 15 GC orbits was conducted based on all available observational data. The orbital parameters given in Table~\ref{TblObs} represent a very robust distribution based on the present position and velocity observations of each cluster. The velocity observations currently have very large uncertainties which can potentially be improved by further proper motion or tidal tail observations. While all clusters in this sample need improvements in their orbital parameters, priority should be given to NGC 1851 and NGC 7078 as these orbits are particularly poorly determined, making the calculation of the stability boundary near useless.

At present the orbital determinations are not quite sufficient to definitely rule out MOND, however NGC 6171 and NGC 6341 have been identified as promising candidates for distinguishing Newtonian and MOND models, requiring only moderate improvement in the observations. In particular, NGC 6171 may already be showing evidence of chaotic diffusion of stars leading to flattening at the chaos radius. Clarification of this phenomenon will require more resolution in the velocity dispersion and better orbital determination.

\section{Acknowledgements}

GFK acknowledges the support by the Chinese Academy of Sciences for young international scientists, grant number O929011001. GFK would also like to thank the anonymous referees whose comments greatly increased the quality of this manuscript. Discussions with Holger Baumgardt during the MODEST 13 meeting in Almaty regarding acceleration in the galactic potential and statistical discussions with James Wicker also improved this work.

\bibliography{GCR2.bib}

\begin{thebibliography}{}

\bibitem[\protect\citeauthoryear{{Aguilar}, {Hut} \& {Ostriker}}{{Aguilar}
  et~al.}{1988}]{AHO1988}
{Aguilar} L.,  {Hut} P.,    {Ostriker} J.~P.,  1988, ApJ, 335, 720

\bibitem[\protect\citeauthoryear{{Allen} \& {Richstone}}{{Allen} \&
  {Richstone}}{1988}]{AR1988}
{Allen} A.~J.,  {Richstone} D.~O.,  1988, ApJ, 325, 583

\bibitem[\protect\citeauthoryear{{Allen}, {Moreno} \& {Pichardo}}{{Allen}
  et~al.}{2006}]{AMP2006}
{Allen} C.,  {Moreno} E.,    {Pichardo} B.,  2006, ApJ, 652, 1150

\bibitem[\protect\citeauthoryear{{Anderson} \& {King}}{{Anderson} \&
  {King}}{2003}]{AndersonKing2003}
{Anderson} J.,  {King} I.~R.,  2003, AJ, 126, 772

\bibitem[\protect\citeauthoryear{{Baumgardt}, {C{\^o}t{\'e}}, {Hilker},
  {Rejkuba}, {Mieske}, {Djorgovski} \& {Stetson}}{{Baumgardt}
  et~al.}{2009}]{BCHRMDS2009}
{Baumgardt} H.,  {C{\^o}t{\'e}} P.,  {Hilker} M.,  {Rejkuba} M.,  {Mieske} S.,
  {Djorgovski} S.~G.,    {Stetson} P.,  2009, MNRAS, 396, 2051

\bibitem[\protect\citeauthoryear{{Baumgardt}, {Grebel} \& {Kroupa}}{{Baumgardt}
  et~al.}{2005}]{BGK2005}
{Baumgardt} H.,  {Grebel} E.~K.,    {Kroupa} P.,  2005, MRNAS, 359, L1

\bibitem[\protect\citeauthoryear{{Baumgardt}, {Hut} \& {Heggie}}{{Baumgardt}
  et~al.}{2002}]{BaumgardtHutHeggie2002}
{Baumgardt} H.,  {Hut} P.,    {Heggie} D.~C.,  2002, MNRAS, 336, 1069

\bibitem[\protect\citeauthoryear{{Baumgardt} \& {Mieske}}{{Baumgardt} \&
  {Mieske}}{2008}]{BM2008}
{Baumgardt} H.,  {Mieske} S.,  2008, MNRAS, 391, 942

\bibitem[\protect\citeauthoryear{{Bedin}, {Piotto}, {King} \&
  {Anderson}}{{Bedin} et~al.}{2003}]{BedinEtAl2003}
{Bedin} L.~R.,  {Piotto} G.,  {King} I.~R.,    {Anderson} J.,  2003, AJ, 126,
  247

\bibitem[\protect\citeauthoryear{{Bekki}}{{Bekki}}{2010}]{Bekki2010}
{Bekki} K.,  2010, ApJL, 724, L99

\bibitem[\protect\citeauthoryear{{Bellazzini}}{{Bellazzini}}{2004}]{Bellazzini%
2004}
{Bellazzini} M.,  2004, MNRAS, 347, 119

\bibitem[\protect\citeauthoryear{{Bellazzini}, {Bragaglia}, {Carretta},
  {Gratton}, {Lucatello}, {Catanzaro} \& {Leone}}{{Bellazzini}
  et~al.}{2012}]{BellazziniEtAl2012}
{Bellazzini} M.,  {Bragaglia} A.,  {Carretta} E.,  {Gratton} R.~G.,
  {Lucatello} S.,  {Catanzaro} G.,    {Leone} F.,  2012, A\&A, 538, A18

\bibitem[\protect\citeauthoryear{{Bianchini}, {Varri}, {Bertin} \&
  {Zocchi}}{{Bianchini} et~al.}{2013}]{BianchiniEtAl2013}
{Bianchini} P.,  {Varri} A.~L.,  {Bertin} G.,    {Zocchi} A.,  2013, ApJ, 772,
  67

\bibitem[\protect\citeauthoryear{{Binney} \& {Tremaine}}{{Binney} \&
  {Tremaine}}{1987}]{BinneyTremaine1987}
{Binney} J.,  {Tremaine} S.,  1987, {Galactic dynamics}.
Princeton, NJ, Princeton University Press, 1987

\bibitem[\protect\citeauthoryear{{Boyles}, {Lorimer}, {Turk}, {Mnatsakanov},
  {Lynch}, {Ransom}, {Freire} \& {Belczynski}}{{Boyles}
  et~al.}{2011}]{BoylesEtAl2011}
{Boyles} J.,  {Lorimer} D.~R.,  {Turk} P.~J.,  {Mnatsakanov} R.,  {Lynch}
  R.~S.,  {Ransom} S.~M.,  {Freire} P.~C.,    {Belczynski} K.,  2011, ApJ, 742,
  51

\bibitem[\protect\citeauthoryear{{Chun}, {Kim}, {Sohn}, {Park}, {Han}, {Kim},
  {Lee}, {Lee}, {Lee} \& {Sohn}}{{Chun} et~al.}{2010}]{ChunEtAl2010}
{Chun} S.-H.,  {Kim} J.-W.,  {Sohn} S.~T.,  {Park} J.-H.,  {Han} W.,  {Kim}
  H.-I.,  {Lee} Y.-W.,  {Lee} M.~G.,  {Lee} S.-G.,    {Sohn} Y.-J.,  2010, AJ,
  139, 606

\bibitem[\protect\citeauthoryear{{Cudworth} \& {Hanson}}{{Cudworth} \&
  {Hanson}}{1993}]{CH1993}
{Cudworth} K.~M.,  {Hanson} R.~B.,  1993, AJ, 105, 168

\bibitem[\protect\citeauthoryear{{Dejonghe}}{{Dejonghe}}{1987}]{Dejonghe1987}
{Dejonghe} H.,  1987, MNRAS, 224, 13

\bibitem[\protect\citeauthoryear{{Di Cecco}, {Zocchi}, {Varri}, {Monelli},
  {Bertin}, {Bono}, {Stetson}, {Nonino}, {Buonanno}, {Ferraro}, {Iannicola},
  {Kunder} \& {Walker}}{{Di Cecco} et~al.}{2013}]{DiCeccoEtAl2013}
{Di Cecco} A.,  {Zocchi} A.,  {Varri} A.~L.,  {Monelli} M.,  {Bertin} G.,
  {Bono} G.,  {Stetson} P.~B.,  {Nonino} M.,  {Buonanno} R.,  {Ferraro} I.,
  {Iannicola} G.,  {Kunder} A.,    {Walker} A.~R.,  2013, AJ, 145, 103

\bibitem[\protect\citeauthoryear{{Di Criscienzo}, {D'Antona}, {Milone},
  {Ventura}, {Caloi}, {Carini}, {D'Ercole}, {Vesperini} \& {Piotto}}{{Di
  Criscienzo} et~al.}{2011}]{ChemNGC2419}
{Di Criscienzo} M.,  {D'Antona} F.,  {Milone} A.~P.,  {Ventura} P.,  {Caloi}
  V.,  {Carini} R.,  {D'Ercole} {Vesperini} E.,    {Piotto} G.,  2011, ArXiv
  e-prints

\bibitem[\protect\citeauthoryear{{Dinescu}, {Girard}, {van Altena}, {Mendez} \&
  {Lopez}}{{Dinescu} et~al.}{1997}]{DinescuI}
{Dinescu} D.~I.,  {Girard} T.~M.,  {van Altena} W.~F.,  {Mendez} R.~A.,
  {Lopez} C.~E.,  1997, AJ, 114, 1014

\bibitem[\protect\citeauthoryear{{Dinescu}, {van Altena}, {Girard} \&
  {L{\'o}pez}}{{Dinescu} et~al.}{1999}]{DinescuII}
{Dinescu} D.~I.,  {van Altena} W.~F.,  {Girard} T.~M.,    {L{\'o}pez} C.~E.,
  1999, AJ, 117, 277

\bibitem[\protect\citeauthoryear{{Drukier}, {Cohn}, {Lugger}, {Slavin},
  {Berrington} \& {Murphy}}{{Drukier} et~al.}{2007}]{DCLSBM2007}
{Drukier} G.~A.,  {Cohn} H.~N.,  {Lugger} P.~M.,  {Slavin} S.~D.,  {Berrington}
  R.~C.,    {Murphy} B.~W.,  2007, AJ, 133, 1041

\bibitem[\protect\citeauthoryear{{Drukier}, {Slavin}, {Cohn}, {Lugger},
  {Berrington}, {Murphy} \& {Seitzer}}{{Drukier} et~al.}{1998}]{DSCLBMS1998}
{Drukier} G.~A.,  {Slavin} S.~D.,  {Cohn} H.~N.,  {Lugger} P.~M.,  {Berrington}
  R.~C.,  {Murphy} B.~W.,    {Seitzer} P.~O.,  1998, AJ, 115, 708

\bibitem[\protect\citeauthoryear{{Einsel} \& {Spurzem}}{{Einsel} \&
  {Spurzem}}{1999}]{EinselSpurzem1999}
{Einsel} C.,  {Spurzem} R.,  1999, MNRAS, 302, 81

\bibitem[\protect\citeauthoryear{{Feigelson} \& {Jogesh Babu}}{{Feigelson} \&
  {Jogesh Babu}}{2012}]{AstroStats2012}
{Feigelson} E.~D.,  {Jogesh Babu} G.,  2012, {Modern Statistical Methods for
  Astronomy}

\bibitem[\protect\citeauthoryear{{Fellhauer}, {Evans}, {Belokurov}, {Wilkinson}
  \& {Gilmore}}{{Fellhauer} et~al.}{2007}]{FellhauerEtAl2007}
{Fellhauer} M.,  {Evans} N.~W.,  {Belokurov} V.,  {Wilkinson} M.~I.,
  {Gilmore} G.,  2007, MNRAS, 380, 749

\bibitem[\protect\citeauthoryear{{Freire}, {Camilo}, {Kramer}, {Lorimer},
  {Lyne}, {Manchester} \& {D'Amico}}{{Freire} et~al.}{2003}]{FreireEtAl2003}
{Freire} P.~C.,  {Camilo} F.,  {Kramer} M.,  {Lorimer} D.~R.,  {Lyne} A.~G.,
  {Manchester} R.~N.,    {D'Amico} N.,  2003, MNRAS, 340, 1359

\bibitem[\protect\citeauthoryear{{Fukushige} \& {Heggie}}{{Fukushige} \&
  {Heggie}}{2000}]{FH2000}
{Fukushige} T.,  {Heggie} D.~C.,  2000, MNRAS, 318, 753

\bibitem[\protect\citeauthoryear{{Geffert}}{{Geffert}}{1998}]{Geffert1998}
{Geffert} M.,  1998, A\&A, 340, 305

\bibitem[\protect\citeauthoryear{{Gentile}, {Famaey}, {Angus} \&
  {Kroupa}}{{Gentile} et~al.}{2010}]{GFAK2010}
{Gentile} G.,  {Famaey} B.,  {Angus} G.,    {Kroupa} P.,  2010, A\&A, 509, A97

\bibitem[\protect\citeauthoryear{Gnedin, Lee \& Ostriker}{Gnedin
  et~al.}{1999}]{GLO1999}
Gnedin O.~Y.,  Lee H.~M.,    Ostriker J.~P.,  1999, ApJ, 522, 935

\bibitem[\protect\citeauthoryear{Gnedin \& Ostriker}{Gnedin \&
  Ostriker}{1997}]{GO1997}
Gnedin O.~Y.,  Ostriker J.~P.,  1997, ApJ, 474, 223

\bibitem[\protect\citeauthoryear{{Gratton}, {Carretta} \&
  {Bragaglia}}{{Gratton} et~al.}{2012}]{GCB2012}
{Gratton} R.~G.,  {Carretta} E.,    {Bragaglia} A.,  2012, A\&ARv, 20, 50

\bibitem[\protect\citeauthoryear{{Grillmair}, {Freeman}, {Irwin} \&
  {Quinn}}{{Grillmair} et~al.}{1995}]{GrillmairEtAl1995}
{Grillmair} C.~J.,  {Freeman} K.~C.,  {Irwin} M.,    {Quinn} P.~J.,  1995, AJ,
  109, 2553

\bibitem[\protect\citeauthoryear{{Guo}}{{Guo}}{1995}]{Guo1995}
{Guo} X.,  1995, PhD thesis, Yale University.

\bibitem[\protect\citeauthoryear{{Haghi}, {Baumgardt} \& {Kroupa}}{{Haghi}
  et~al.}{2011}]{HBK2011}
{Haghi} H.,  {Baumgardt} H.,    {Kroupa} P.,  2011, A\&A, 527, A33

\bibitem[\protect\citeauthoryear{{Harris}}{{Harris}}{1996}]{Harris1996}
{Harris} W.~E.,  1996, AJ, 112, 1487

\bibitem[\protect\citeauthoryear{{Ibata}, {Nipoti}, {Sollima}, {Bellazzini},
  {Chapman} \& {Dalessandro}}{{Ibata} et~al.}{2013}]{IbataEtAl2013}
{Ibata} R.,  {Nipoti} C.,  {Sollima} A.,  {Bellazzini} M.,  {Chapman} S.~C.,
  {Dalessandro} E.,  2013, MNRAS, 428, 3648

\bibitem[\protect\citeauthoryear{{Ibata}, {Sollima}, {Nipoti}, {Bellazzini},
  {Chapman} \& {Dalessandro}}{{Ibata} et~al.}{2011}]{IbataEtAl2011}
{Ibata} R.,  {Sollima} A.,  {Nipoti} C.,  {Bellazzini} M.,  {Chapman} S.~C.,
  {Dalessandro} E.,  2011, ArXiv e-prints

\bibitem[\protect\citeauthoryear{{Jeffreys}}{{Jeffreys}}{1998}]{Jeffreys1998}
{Jeffreys} H.,  1998, {The Theory of Probability}.
Oxford University Press, 3rd Ed. 1998

\bibitem[\protect\citeauthoryear{{Jordi} \& {Grebel}}{{Jordi} \&
  {Grebel}}{2010}]{JordiGrebel2010}
{Jordi} K.,  {Grebel} E.~K.,  2010, A\&A, 522, A71

\bibitem[\protect\citeauthoryear{{Kalirai}, {Richer}, {Hansen}, {Stetson},
  {Shara}, {Saviane}, {Rich}, {Limongi}, {Ibata}, {Gibson}, {Fahlman} \&
  {Brewer}}{{Kalirai} et~al.}{2004}]{KaliraiEtAl2004}
{Kalirai} J.~S.,  {Richer} H.~B.,  {Hansen} B.~M.,  {Stetson} P.~B.,  {Shara}
  M.~M.,  {Saviane} I.,  {Rich} R.~M.,  {Limongi} M.,  {Ibata} R.,  {Gibson}
  B.~K.,  {Fahlman} G.~G.,    {Brewer} J.,  2004, ApJ, 601, 277

\bibitem[\protect\citeauthoryear{{Keenan}}{{Keenan}}{1981}]{Keenan1981}
{Keenan} D.~W.,  1981, A\&A, 95, 340

\bibitem[\protect\citeauthoryear{{Kennedy}}{{Kennedy}}{2014}]{PaperI}
{Kennedy} G.~F.,  2014, MNRAS, 444, 3328

\bibitem[\protect\citeauthoryear{King}{King}{1962}]{KingI}
King I.,  1962, AJ, 67, 471

\bibitem[\protect\citeauthoryear{{Kouwenhoven} \& {de Grijs}}{{Kouwenhoven} \&
  {de Grijs}}{2009}]{ThijsRichard2009}
{Kouwenhoven} M.~B.~N.,  {de Grijs} R.,  2009, A\&SS, 324, 171

\bibitem[\protect\citeauthoryear{{Kundic} \& {Ostriker}}{{Kundic} \&
  {Ostriker}}{1995}]{KundicOstriker1995}
{Kundic} T.,  {Ostriker} J.~P.,  1995, ApJ, 438, 702

\bibitem[\protect\citeauthoryear{{K{\"u}pper}, {Kroupa}, {Baumgardt} \&
  {Heggie}}{{K{\"u}pper} et~al.}{2010}]{KupperEtAl2010}
{K{\"u}pper} A.~H.~W.,  {Kroupa} P.,  {Baumgardt} H.,    {Heggie} D.~C.,  2010,
  MNRAS, 401, 105

\bibitem[\protect\citeauthoryear{{Lane}, {Kiss}, {Lewis}, {Ibata}, {Siebert},
  {Bedding} \& {Sz{\'e}kely}}{{Lane} et~al.}{2010}]{LaneEtAl2010}
{Lane} R.~R.,  {Kiss} L.~L.,  {Lewis} G.~F.,  {Ibata} R.~A.,  {Siebert} A.,
  {Bedding} T.~R.,    {Sz{\'e}kely} P.,  2010, MNRAS, 401, 2521

\bibitem[\protect\citeauthoryear{{Lane}, {Kiss}, {Lewis}, {Ibata}, {Siebert},
  {Bedding}, {Sz{\'e}kely} \& {Szab{\'o}}}{{Lane} et~al.}{2011}]{LaneEtAl2011}
{Lane} R.~R.,  {Kiss} L.~L.,  {Lewis} G.~F.,  {Ibata} R.~A.,  {Siebert} A.,
  {Bedding} T.~R.,  {Sz{\'e}kely} P.,    {Szab{\'o}} G.~M.,  2011, A\&A, 530,
  A31

\bibitem[\protect\citeauthoryear{{Lehmann} \& {Scholz}}{{Lehmann} \&
  {Scholz}}{1997}]{LehmannScholz1997}
{Lehmann} I.,  {Scholz} R.-D.,  1997, A\&A, 320, 776

\bibitem[\protect\citeauthoryear{{Leon}, {Meylan} \& {Combes}}{{Leon}
  et~al.}{2000}]{LeonEtAl2000}
{Leon} S.,  {Meylan} G.,    {Combes} F.,  2000, A\&A, 359, 907

\bibitem[\protect\citeauthoryear{{Mardling}}{{Mardling}}{2008}]{RoIoA}
{Mardling} R.~A.,  2008, in {Aarseth} S.~J.,  {Tout} C.~A.,   {Mardling} R.~A.,
   eds, Lecture Notes in Physics, Vol. 760: The Cambridge N-body Lectures
  {Resonance, chaos and stability: the three-body problem in astrophysics}

\bibitem[\protect\citeauthoryear{{Mardling}}{{Mardling}}{2013}]{RoNew}
{Mardling} R.~A.,  2013, ArXiv e-prints

\bibitem[\protect\citeauthoryear{{Marks} \& {Kroupa}}{{Marks} \&
  {Kroupa}}{2010}]{MK2010}
{Marks} M.,  {Kroupa} P.,  2010, MNRAS, 406, 2000

\bibitem[\protect\citeauthoryear{{McLaughlin} \& {Fall}}{{McLaughlin} \&
  {Fall}}{2008}]{McLFall2008}
{McLaughlin} D.~E.,  {Fall} S.~M.,  2008, ApJ, 679, 1272

\bibitem[\protect\citeauthoryear{{McLaughlin} \& {van der Marel}}{{McLaughlin}
  \& {van der Marel}}{2005a}]{McLaughlinVDMarel2005}
{McLaughlin} D.~E.,  {van der Marel} R.~P.,  2005a, ApJS, 161, 304

\bibitem[\protect\citeauthoryear{{McLaughlin} \& {van der Marel}}{{McLaughlin}
  \& {van der Marel}}{2005b}]{McLvdM2005}
{McLaughlin} D.~E.,  {van der Marel} R.~P.,  2005b, ApJS, 161, 304

\bibitem[\protect\citeauthoryear{{McNamara}, {Harrison} \&
  {Baumgardt}}{{McNamara} et~al.}{2004}]{MHB2004}
{McNamara} B.~J.,  {Harrison} T.~E.,    {Baumgardt} H.,  2004, ApJ, 602, 264

\bibitem[\protect\citeauthoryear{{Meziane} \& {Colin}}{{Meziane} \&
  {Colin}}{1996}]{Meziane96}
{Meziane} K.,  {Colin} J.,  1996, A\&A, 306, 747

\bibitem[\protect\citeauthoryear{{Milgrom}}{{Milgrom}}{1983}]{Milgrom1983}
{Milgrom} M.,  1983, ApJ, 270, 365

\bibitem[\protect\citeauthoryear{{Miocchi}, {Lanzoni}, {Ferraro},
  {Dalessandro}, {Vesperini}, {Pasquato}, {Beccari}, {Pallanca} \&
  {Sanna}}{{Miocchi} et~al.}{2013}]{MiocchiEtAl2013}
{Miocchi} P.,  {Lanzoni} B.,  {Ferraro} F.~R.,  {Dalessandro} E.,  {Vesperini}
  E.,  {Pasquato} M.,  {Beccari} G.,  {Pallanca} C.,    {Sanna} N.,  2013, ApJ,
  774, 151

\bibitem[\protect\citeauthoryear{{Miyamoto} \& {Nagai}}{{Miyamoto} \&
  {Nagai}}{1975}]{MiyamotoNagai1975}
{Miyamoto} M.,  {Nagai} R.,  1975, PASJ, 27, 533

\bibitem[\protect\citeauthoryear{{Moffat} \& {Toth}}{{Moffat} \&
  {Toth}}{2008}]{MT2008}
{Moffat} J.~W.,  {Toth} V.~T.,  2008, ApJ, 680, 1158

\bibitem[\protect\citeauthoryear{{Montuori}, {Capuzzo-Dolcetta}, {Di Matteo},
  {Lepinette} \& {Miocchi}}{{Montuori} et~al.}{2007}]{MontuoriEtAl2007}
{Montuori} M.,  {Capuzzo-Dolcetta} R.,  {Di Matteo} P.,  {Lepinette} A.,
  {Miocchi} P.,  2007, ApJ, 659, 1212

\bibitem[\protect\citeauthoryear{{Nipoti}, {Ciotti} \& {Londrillo}}{{Nipoti}
  et~al.}{2011}]{NCL2011}
{Nipoti} C.,  {Ciotti} L.,    {Londrillo} P.,  2011, MNRAS, p.~608

\bibitem[\protect\citeauthoryear{{Odenkirchen}, {Brosche}, {Geffert} \&
  {Tucholke}}{{Odenkirchen} et~al.}{1997}]{Odenkirchen1997}
{Odenkirchen} M.,  {Brosche} P.,  {Geffert} M.,    {Tucholke} H.-J.,  1997, New
  Astronomy, 2, 477

\bibitem[\protect\citeauthoryear{{Ostriker}, {Binney} \& {Saha}}{{Ostriker}
  et~al.}{1989}]{OBS1989}
{Ostriker} J.~P.,  {Binney} J.,    {Saha} P.,  1989, MNRAS, 241, 849

\bibitem[\protect\citeauthoryear{{Ostriker}, {Spitzer} Jr. \&
  {Chevalier}}{{Ostriker} et~al.}{1972}]{OstrikerEtAl1972}
{Ostriker} J.~P.,  {Spitzer} Jr. L.,    {Chevalier} R.~A.,  1972, ApJL, 176,
  L51

\bibitem[\protect\citeauthoryear{Press, Flannery, Teukolsky \& Vettering}{Press
  et~al.}{1986}]{NR86}
Press W.~B.,  Flannery B.~P.,  Teukolsky S.~A.,    Vettering W.~T.,  1986,
  Numerical Recipes: The Art of Scientific Computing.
Cambridge Univ. Press, Cambridge

\bibitem[\protect\citeauthoryear{{Scarpa} \& {Falomo}}{{Scarpa} \&
  {Falomo}}{2010}]{SF2010}
{Scarpa} R.,  {Falomo} R.,  2010, A\&A, 523, A43

\bibitem[\protect\citeauthoryear{{Scarpa}, {Marconi}, {Carraro}, {Falomo} \&
  {Villanova}}{{Scarpa} et~al.}{2011}]{SMCF2011}
{Scarpa} R.,  {Marconi} G.,  {Carraro} G.,  {Falomo} R.,    {Villanova} S.,
  2011, A\&A, 525, A148

\bibitem[\protect\citeauthoryear{{Scarpa}, {Marconi} \& {Gilmozzi}}{{Scarpa}
  et~al.}{2003}]{ScarpaWCen03}
{Scarpa} R.,  {Marconi} G.,    {Gilmozzi} R.,  2003, A\&AL, 405, L15

\bibitem[\protect\citeauthoryear{{Scarpa}, {Marconi} \& {Gilmozzi}}{{Scarpa}
  et~al.}{2004}]{SMG2004}
{Scarpa} R.,  {Marconi} G.,    {Gilmozzi} R.,  2004, in {R.~Dettmar, U.~Klein,
  \& P.~Salucci} ed., Baryons in Dark Matter Halos {Using globular clusters to
  test gravity in the weak acceleration regime: NGC 6171}

\bibitem[\protect\citeauthoryear{{Scarpa}, {Marconi}, {Gilmozzi} \&
  {Carraro}}{{Scarpa} et~al.}{2007a}]{SMGC2007}
{Scarpa} R.,  {Marconi} G.,  {Gilmozzi} R.,    {Carraro} G.,  2007a, The
  Messenger, 128, 41

\bibitem[\protect\citeauthoryear{{Scarpa}, {Marconi}, {Gilmozzi} \&
  {Carraro}}{{Scarpa} et~al.}{2007b}]{ScarpaEtAl2007}
{Scarpa} R.,  {Marconi} G.,  {Gilmozzi} R.,    {Carraro} G.,  2007b, A\&A, 462,
  L9

\bibitem[\protect\citeauthoryear{{Sollima} \& {Nipoti}}{{Sollima} \&
  {Nipoti}}{2010}]{SN2010}
{Sollima} A.,  {Nipoti} C.,  2010, MNRAS, 401, 131

\bibitem[\protect\citeauthoryear{{Testa}, {Zaggia}, {Andreon}, {Longo},
  {Scaramella}, {Djorgovski} \& {de Carvalho}}{{Testa}
  et~al.}{2000}]{Testa2000}
{Testa} V.,  {Zaggia} S.~R.,  {Andreon} S.,  {Longo} G.,  {Scaramella} R.,
  {Djorgovski} S.~G.,    {de Carvalho} R.,  2000, A\&A, 356, 127

\bibitem[\protect\citeauthoryear{{Trager}, {King} \& {Djorgovski}}{{Trager}
  et~al.}{1995}]{TragerEtAl1995}
{Trager} S.~C.,  {King} I.~R.,    {Djorgovski} S.,  1995, AJ, 109, 218

\bibitem[\protect\citeauthoryear{{Vande Putte} \& {Cropper}}{{Vande Putte} \&
  {Cropper}}{2009}]{VPC2009}
{Vande Putte} D.,  {Cropper} M.,  2009, MNRAS, 392, 113

\bibitem[\protect\citeauthoryear{{Webb}, {Harris}, {Sills} \& {Hurley}}{{Webb}
  et~al.}{2013}]{WebbEtAl2013}
{Webb} J.~J.,  {Harris} W.~E.,  {Sills} A.,    {Hurley} J.~R.,  2013, ApJ, 764,
  124

\end{thebibliography}

\end{document}